\documentclass[onecolumn]{revtex4}
\usepackage{graphicx}
\usepackage{enumerate}
\usepackage{amssymb}
\usepackage{amsmath}
\usepackage{amsfonts}
\usepackage{dcolumn}
\usepackage{bm}
\usepackage{framed}
\usepackage{braket}
\usepackage{color}
\usepackage{alphalph}
\usepackage{fancyhdr}
\begin{document}
\title{Electric control of spin orbit coupling in graphene-based nanostructures with broken rotational symmetry}

\author{A. Ciattoni$^1$}
\email{alessandro.ciattoni@aquila.infn.it}
\author{C. Conti$^{2}$}
\author{A. V. Zayats$^4$}
\author{A. Marini$^5$}
\email{andrea.marini@aquila.infn.it}
\affiliation{$^1$CNR-SPIN, c/o Dip.to di Scienze Fisiche e Chimiche, Via Vetoio, 67100 Coppito (L'Aquila), Italy}
\affiliation{$^2$Institute for Complex Systems, National Research Council (ISC-CNR), Via dei Taurini 19, 00185, Rome, Italy}
\affiliation{$^4$Department of Physics and London Centre for Nanotechnology, King’s College London, Strand, London, WC2R 2LS, United Kingdom }
\affiliation{$^5$Department of Physical and Chemical Sciences, University of L'Aquila, Via Vetoio, 67100 L'Aquila, Italy}
\date{\today}


\begin{abstract} 
Spin and angular momenta of light are important degrees of freedom in nanophotonics which control light propagation, optical forces and information encoding. Typically, optical angular momentum is generated using q-plates or spatial light modulators. Here, we show that graphene-supported plasmonic nanostructures with broken rotational symmetry provide a surprising spin to orbital angular momentum conversion, which can be continuously controlled by changing the electrochemical potential of graphene. Upon resonant illumination by a circularly polarized plane wave, a polygonal array of indium-tin-oxide nanoparticles on a graphene sheet generates scattered field carrying electrically-tunable orbital angular momentum. This unique photonic spin-orbit coupling occurs due to the strong coupling of graphene plasmon polaritons and localised surface plasmons of the nanoparticles and leads to the controlled directional excitation of graphene plasmons. The tuneable spin-orbit conversion pave the way to high-rate information encoding in optical communications, electric steering functionalities in optical tweezers, and nanorouting of higher-dimensional entangled photon states.
\end{abstract}

\maketitle
The uncertainty principle rules the localization of particles and has fundamental implications in many new technological developments. The large momentum components of a strongly localized wave-packet entail complex field dynamics absent in the case of a weak confinement. Many physical phenomena accompany sub-wavelength localization of photons, including the photonic spin-orbit interaction (SOI) between the polarization and spatial degrees of freedom of light which are governed by the spin angular momentum (SAM) and orbital angular momentum (OAM)~\cite{Barne1,VanEn1,Bliok2}. In the paraxial regime (weak localization), SOI is usually negligible~\cite{Allen1} but may be augmented by inhomogeneities \cite{Onoda1,Hoste1,Bliok1}, anisotropy \cite{Ciatt1,Brass1}, and epsilon-near-zero materials \cite{Ciatt2,lsa-16}. Non-paraxial focusing~\cite{Zhaoo1,Niemi1} by small particles~\cite{Bliok3,Dogar1} or a doped graphene~\cite{Ciatt3} enhances SOI and results in spin-dependent directionalities and optical vortices \cite{Bliok2}. In the case of the extreme confinement at the subwavelength scales, SOI is dominant~\cite{Rodri1,Neuge1,Peter1,Sukho1} with applications in quantum optics \cite{Luxmo1,Mitsc1}, high-resolution microscopy \cite{Rodri3,nat-phys}, beam shaping with planar metasurfaces \cite{Yuuuu1}, optical forces \cite{Rodri2,prb-18,nl-18} and nanophotonics \cite{Espin1,Rodri4}.

The physical origin of the strong SOI in the highly confined electromagnetic fields~\cite{Bliok4,Kimmm1,Bliok5,Neuge2,Beksh1,Antog1} are the large longitudinal components of the evanescent waves that produce a SAM orthogonal to the direction of propagation (transverse SAM). Under the reversal of the direction of the wave-vector, the transverse SAM changes sign and the spin-momentum locking \cite{Meche1,Kalho1,Triol1} can be regarded as an analogue of the quantum spin Hall effect (QSHE)~\cite{Bliok6}. QSHE has been observed with surface plasmon polaritons at metal/dielectric interfaces \cite{Muell1,OConn1,Rodri2,Pannn1}, guided waves, graphene plasmon polaritons (GPPs) \cite{Wangg1} and structured guided waves \cite{arxiv-19}.

SAM to OAM conversion is among the most important SOI phenomena and enables optical micromanipulation \cite{Fries1}, entanglement \cite{Mairr1,Fickl1} and quantum protocols with higher-dimensional quantum states \cite{Vazir1}. Many strategies have been proposed for OAM generation, including anisotropic media \cite{Ciatt0}, inhomogeneous liquid crystals (q-plates) \cite{Marru1}, metasurfaces \cite{Karimi} and whispering-gallery mode resonators \cite{Shaoo1}. All these methods operate in the paraxial regime and generate macroscopic optical vortices with defined topological charge, ruling out the possibility of a continuous tuning of OAM with the external control signal (tuneability has only been achieved by changing the polarization state of the input optical beam). OAM to SAM conversion can take place on the nanoscale leading to subwavelength features of spin-skyrmions \cite{nat-phys}. SAM to OAM conversion is, however, difficult to achieve at the nanoscale as it requires spatial manipulation of the beam on subwavelength scales, and the external, continuous control of SOI is practically unexplored. At the same time, the fast encoding of OAM states is important for the development of its use in optical communications \cite{wilner}.

Here, we describe a novel optical SOI mechanism in a hybrid system of a graphene-supported plasmonic nanostructure with broken rotational invariance and report electrically tunable SAM to OAM conversion at the nanoscale, exploiting graphene doping by external electric bias. We also show that this process is accompanied by the emission of confined grahene plasmon polaritons, which can also be electrically controlled. We consider a polygonal array of doped indium-tin-oxide nanospheres deposited on top of a graphene sheet that--once illuminated by a circularly polarized plane wave--generates the scattered field carrying OAM. This OAM can be continuously tuned by doping graphene under the external bias which results in the modification of coupling between localised surface plasmons (LSPs) of the nanoparticles and GPPs. Therefore, ultrafast electrical modulation of OAM can be achieved. The coupling between LSPs and GPPs leads to the precession of the spin of a circular dipole moment of the nanoparticles resulting in the polarization ellipses out of the graphene plane. Due to the tilted rotation axis, the dipoles excite spin-momentum locked GPPs. The near-field interference of these circulating plasmonic fields produces a non-vanishing OAM. Changing the graphene Fermi energy tunes the value of the generated OAM, because of the spectral shift of the graphene plasmonic resonance and the resulting modification of the coupling between GPPs and LSPs.

\begin{figure*} \label{Fig1}
\center
\includegraphics[width=1\textwidth]{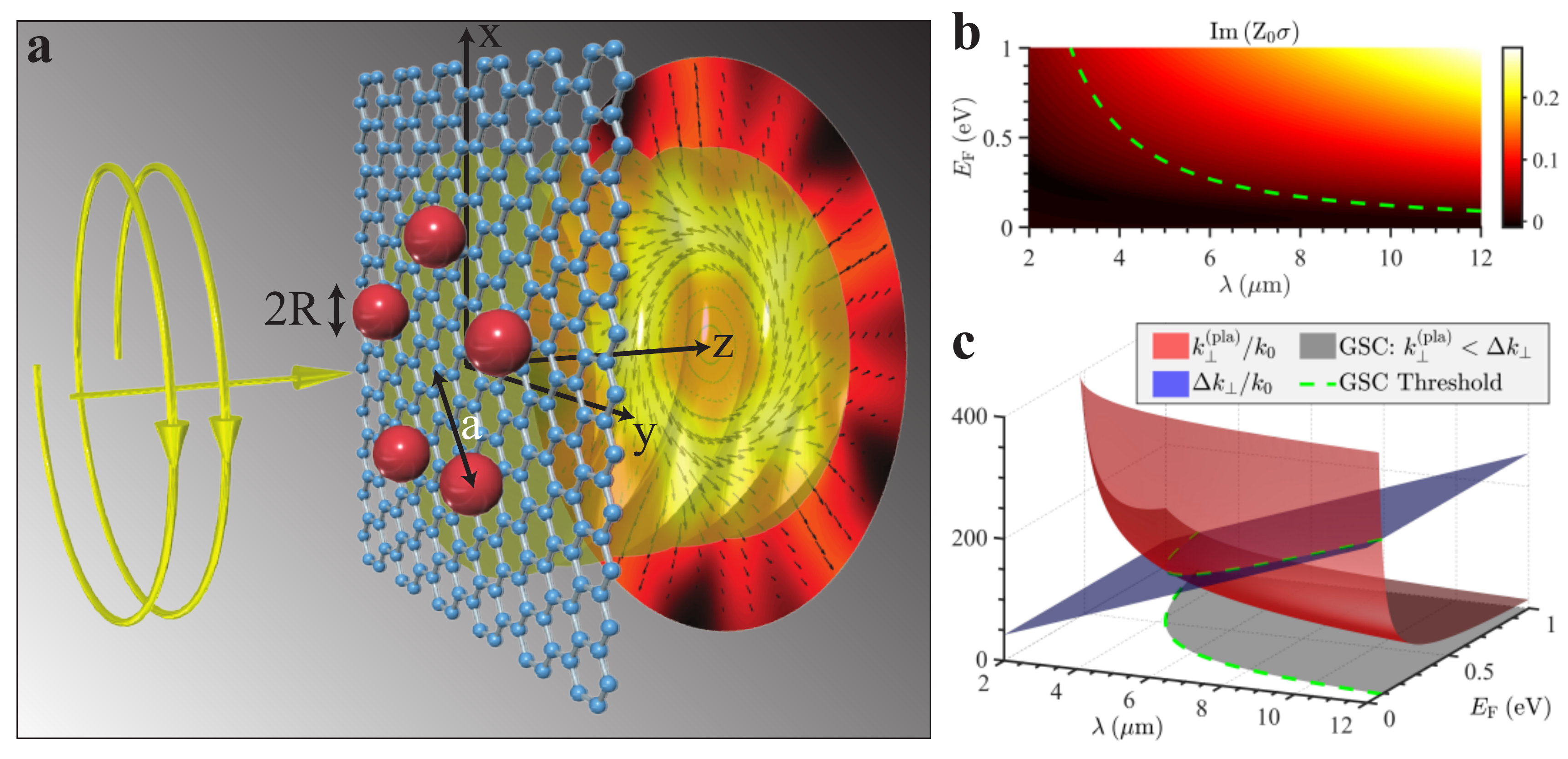}
\caption{{\bf Graphene-nanoparticle system for electrically tunable SAM to OAM conversion.} {\bf a} Schematics of the system: the ITO nanospheres of radius $R$ positioned at the vertices of a regular polygon of radius $a$ on a graphene sheet. A normally incident circularly polarized plane wave feeds the nanostructure with SAM. The scattered field (whose amplitude is sketched in a sample situation as the colour plot) carries highly tunable OAM and the linear momentum density distribution (black arrows) swirls around the polygon axis. {\bf b,c} The dependences of (b) the imaginary part of the normalized graphene surface conductivity $\sigma$ ($Z_0$ is the vacuum impedance) and (c) GPP wavenumber $k_ \bot ^{\left( {\rm pla} \right)}$ (red surface) and the  spatial spectral width $\Delta k_\bot$ of the reflected evanescent field mediating the LSP-GPP coupling (blue surface) on the wavelength $\lambda$ and the graphene Fermi energy $E_{\rm F}$. The green dashed curves in b and c indicate the GSC threshold when the spectrum of  the field generated by the nanoparticles and reflected from the graphene sheet contains the GPP wavenumber $k_ \bot ^{\left( {\rm pla} \right)}  < \Delta k_\bot$ (grey region).}
\end{figure*}

\begin{figure*} \label{Fig2}
\center
\includegraphics[width=1\textwidth]{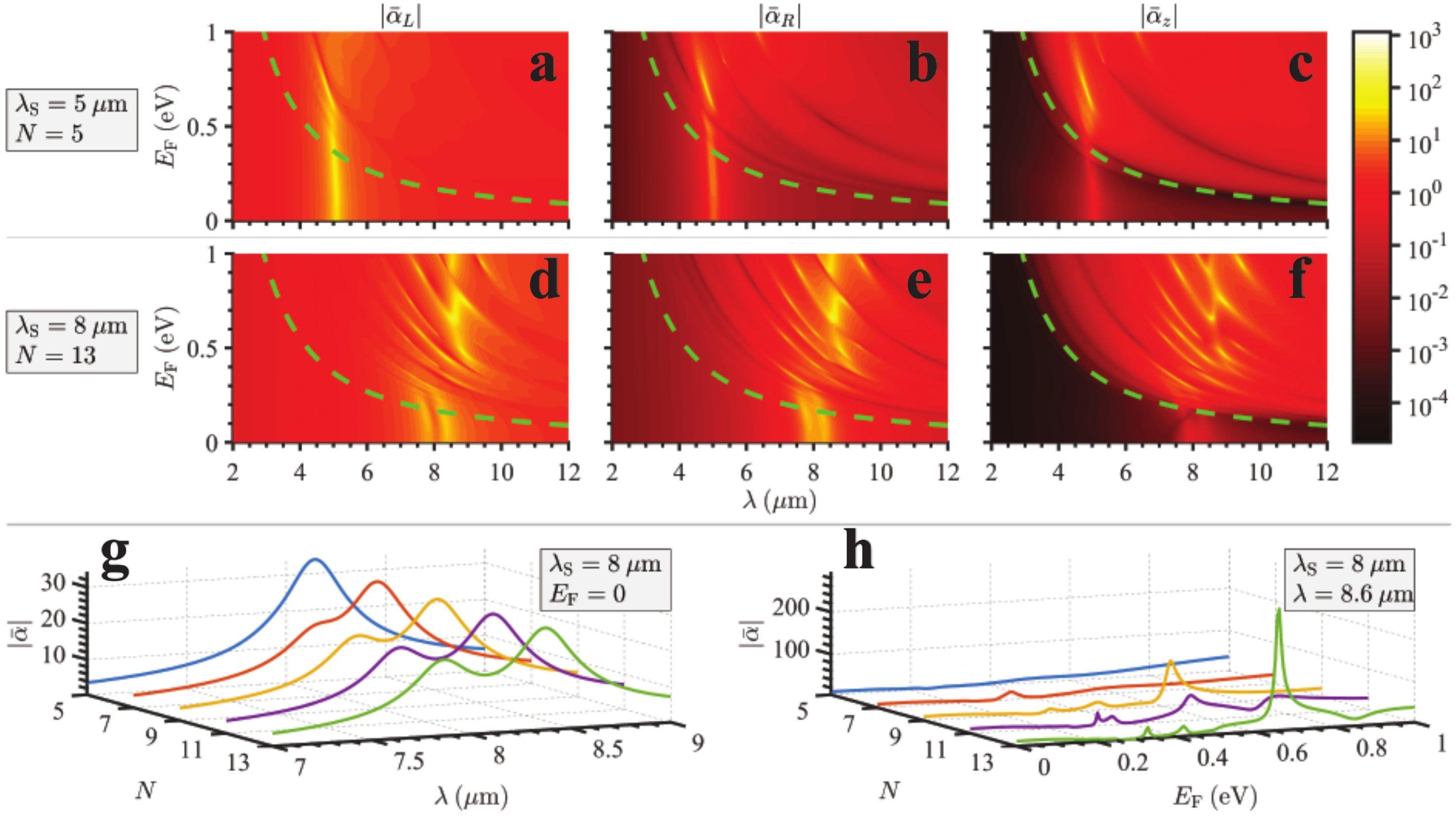}
\caption{{\bf Dipole excitations and hybrid plasmonic resonances.} {\bf a-f} Dimensionless generalized polarizability amplitudes of a polygon with {\bf a-c} $N=5$ and {\bf d-f} $N=13$ ITO nanoparticles resonating at {\bf a-c} $\lambda_S = 5 \: \mu$m and {\bf d-f} $\lambda_S = 8 \: \mu$m. The green dashed lines indicate the GSC threshold as in Fig.~1.
  Below the GSC threshold, the longitudinal polarizabililty ${\bar \alpha}_z$ is always negligible,
  whereas it is comparable with the other components ${\bar \alpha}_{L,R}$ above the GSC threshold due to graphene plasmon excitation. {\bf g,h} Dependences of $\left| {\bar \alpha } \right|$ for {\bf g} $\lambda_S = 8 \: \mu$m and $E_{\rm F} =0$ (nonresonant graphene) on the wavelength $\lambda$ and {\bf h} $\lambda = 8.6 \: \mu$m on the Fermi energy $E_{\rm F}$ for several nanosphere numbers $N$.}
\end{figure*}

{\bf Results}

{\bf Hybrid plasmonic resonances.} We begin by analyzing the collective dynamics of a polygonal array of $N$ nanoparticles placed near an extended graphene sheet illuminated by a plane wave carrying only SAM (Fig.1{\bf a}). The system is not invariant under the full group of rotations around the $z$-axis, but only with respect to the sub-group $C_N$ of discrete rotations of angles $2\pi r/N$.
Therefore, the $z$-component of the radiation angular momentum is not conserved in scattering processes. The graphene sheet lies on the plane $z=0$ between the half spaces $z<0$ and $z>0$ which are filled by two transparent dielectrics of permittivities $\varepsilon_1$ and $\varepsilon_2$, respectively. The nanoparticles of radius $R$ are arranged at the vertices of a regular polygon of radius $a$
on the graphene sheet. The nanoparticle centers are at ${\bf{r}}^{\left( q \right)}  = a\left[ {\cos \left( 2\pi q /N  \right){\bf{e}}_x  + \sin \left( 2\pi q /N  \right){\bf{e}}_y } \right] - R{\bf{e}}_z$ ($q=0,...,N-1$). We here consider indium-tin-oxide nanoparticles of permittivity $\varepsilon_{\rm S} (\lambda)$. The nanoparticles have a localised plasmon resonance at a target infrared wavelength $\lambda_{\rm S}$ which can be controlled by the doping. The infrared response of graphene is governed by the wavelength-dependent surface conductivity $\sigma(\lambda)$ (Fig.~1{\bf b}), which we calculate in the random phase approximation (RPA) \cite{Wunsch}. The incident radiation ${\bf E}_0 = E_0 {\bf e}_{\rm L} e^{ik_1 z} $ is a monochromatic ($e^{-i \omega t}$), left circularly polarized plane wave, with $k_1 = \sqrt{\varepsilon_1} \omega /c$ and ${\bf e}_{\rm L} = ({\bf e}_x + i {\bf e}_y) / \sqrt{2}$.

Since the infrared wavelengths are much larger than the particles radius, the $q-$th nanosphere responds with an induced dipole ${\bf p}^{(q)} = \alpha {\bf E}^{(\rm ext)} ({\bf r}^{(q)})$, where $\alpha  = 4\pi \varepsilon _0 \varepsilon _1 R^3 \left( {\frac{{\varepsilon _{\rm S}  - \varepsilon _1 }}{{\varepsilon _{\rm S}  + 2\varepsilon _1 }}} \right)$ is its electric polarizability and ${\bf E}^{(\rm ext)} ({\bf r}^{(q)})$ is the external field at the nanoparticle center.
We calculate multiple scattering of all orders by generalizing the discrete dipole model \cite{Chaum1} to account for the interaction between the nanoparticles and the effect of a graphene substrate. The field ${\bf E}^{(\rm ext)} ({\bf r}^{(q)})$ contains the incident field, the fields of all the other $N-1$ dipoles, and the total field reflected by graphene. The dipole moments are then determined by
\begin{equation} \label{CDM}
{\frac{1}{\alpha }{\bf{p}}^{\left( q \right)}  - \sum\limits_{s = 0}^{N - 1} {\tilde H^{\left( {q,s} \right)} } {\bf{p}}^{\left( s \right)}  = \left[ {e^{ik_1 R} r_{\rm TE} \left( 0 \right) + e^{ - ik_1 R} } \right]E_0 {\bf{e}}_{\rm L} },
\end{equation}
where $\tilde H^{\left( {q,s} \right)}$ is the interaction matrix and $r_{\rm TE} \left( 0 \right)$ is the reflectivity of the graphene sheet for the normally incident (transverse-electric) plane waves (details in the  Supplementary Information).
Using the dipole moments from Eq.(\ref{CDM}), we find the scattered field from its angular spectrum representation. The $3N$ hybrid plasmonic resonances of the graphene-nanoparticles system, corresponding to the $3N$ components of the coupled nanoparticle dipoles, can be found from Eq.~(\ref{CDM}), assuming the incident plane wave is absent ($E_0 = 0$). This hybridization is due to the coupling between nanoparticle LSPs and GSPs. The spatial spectrum of the reflected field is dominated by the evanescent waves with an asymptotic tail $e^{-2 k_\bot R}$, where $2R$ is the total distance traveled by the field from the polygon plane ($z=-R$) to graphene ($z=0$) and back. The coupling to GPP occurs when the spectrum of the reflected field scattered by the nanoparticles contains a GPP wavenumber $k_ \bot ^{\left( {\rm pla} \right)}  < \Delta k_\bot$ (Fig. 1{\bf c}).

The invariance of the graphene-nanoparticles system under the group $C_N$ implies that the $3N$ dimensional dipoles space breaks into $N$ tri-dimensional invariant subspaces each carrying an irreducible representation of the group (see Supplementary Information). The impinging plane wave $E_0 {\bf{e}}_L$ selects only one of these invariant subspaces that is characterized by the appropriate relation between the dipole moments ($q=0,...,N$):
\begin{equation} \label{invar}
{\bf{p}}^{\left( q \right)}  = e^{i\frac{{2\pi }}{N}q} R_q {\bf{p}}^{\left( 0 \right)},
\end{equation}
where $R_q$ is the dyadic operator which rotates a vector around the $z$-axis by an angle $2 \pi q/N$. The $q-$th dipole coincides with the $0-$th dipole rotated by the angle $2 \pi q /N$, multiplied by the phase factor $e^{i\frac{{2\pi }}{N}q}$, in accordance with the Bloch theorem for discrete rotational symmetry. This symmetry enables to cast Eq.(\ref{CDM}) as a single equation for the $0-$th dipole moment
\begin{equation}
{\bf{p}}^{\left( 0 \right)}  \equiv 4\pi \varepsilon _0 R^3 \left( {\bar \alpha _L {\bf{e}}_L  + \bar \alpha _R {\bf{e}}_R  + \bar \alpha _z {\bf{e}}_z } \right)E_0,
\end{equation}
where ${\bf{e}}_R = {\bf{e}}_L^*$ is the right-hand circular unit vector and $\bar \alpha _L$, $\bar \alpha _R$ and $\bar \alpha _z$ are the dimensionless generalized dipole polarizabilities in the circular basis.

In the numerical calculations, we consider $\varepsilon_1 = 2.0136$, $\varepsilon_2 = 2.0851$, $R = 30$ nm and $a = \frac{{3R}}{{2\sin \left( {\pi /N} \right)}}$ corresponding to the polygon edge equal to $3R$. We model the ITO nanospheres permittivity with the Drude model $\varepsilon _{\rm S} \left( \omega  \right) = 1 - \frac{{\omega _{\rm p}^2 }}{{\omega ^2  + i\omega \Gamma }}$.
The plasma frequency $\omega_{\rm p}$ can be tuned in such a way that the particles resonate at the vacuum wavelength $\lambda_S = 2\pi c / \omega_{\rm S}$ (i.e. $\varepsilon _{\rm S} \left( {\omega _{\rm S }} \right) =  - 2\varepsilon _1  + 0.1 i$) in order to achieve strong coupling.

The evolution of the hybrid plasmonic resonances with the graphene Fermi energy shows a clear graphene strong coupling (GSC) threshold (Fig.~2). Below the threshold, graphene is not resonant and the dipole excitation is not affected by the weak field reflected from a graphene. In this region, the longitudinal polarizability $\bar \alpha_z$ is negligible because the dipole-dipole interaction fields are in plane with the graphene sheet. Furthermore, the $R$ polarizability $\bar \alpha_R$ produced by the dipole-dipole direct interaction between the nanoparticles is also small, because the impinging $L$ field only excites the $L$ polarizability $\bar \alpha_L$ of the nanoparticles. When the wavelength is close to $\lambda_S$, both $\bar \alpha_L$ and $\bar \alpha_R$ are large owing to the LSP excitation.
In this case, the hybridization due to the dipole-dipole interaction in the nanoparticle polygon produces the splitting of the LSP resonance into two peaks, clearly observed when the distance betwen the nanoparticles become smaller as for N=13 polygon (Fig.~2d,e). The impact of graphene on the dipole excitation process dramatically increases above the GSC threshold (Fig. 2 {\bf a}-{\bf f}). In this regime, a significant longitudinal polarizability $\bar \alpha_z$ arises from the excitation of GPPs with giant electric field $z-$ component, which orients the dipoles in the longitudinal direction. Resonance hybridization is particularly emphasized for the wavelengths close to the nanoparticle LSP wavelength $\lambda_S$. The dependence of the hybrid resonances on the number of nanoparticles in the polygon shows that as $N$ increases, the dipole-dipole interaction splitting of the nanoparticle resonance increases, as the distance between the nanoparticles decreases (Fig. 2{\bf g}). The quality factors of hybridized modes increase with the number of nanoparticles.

\begin{figure*} \label{Fig3}
\center
\includegraphics[width=1\textwidth]{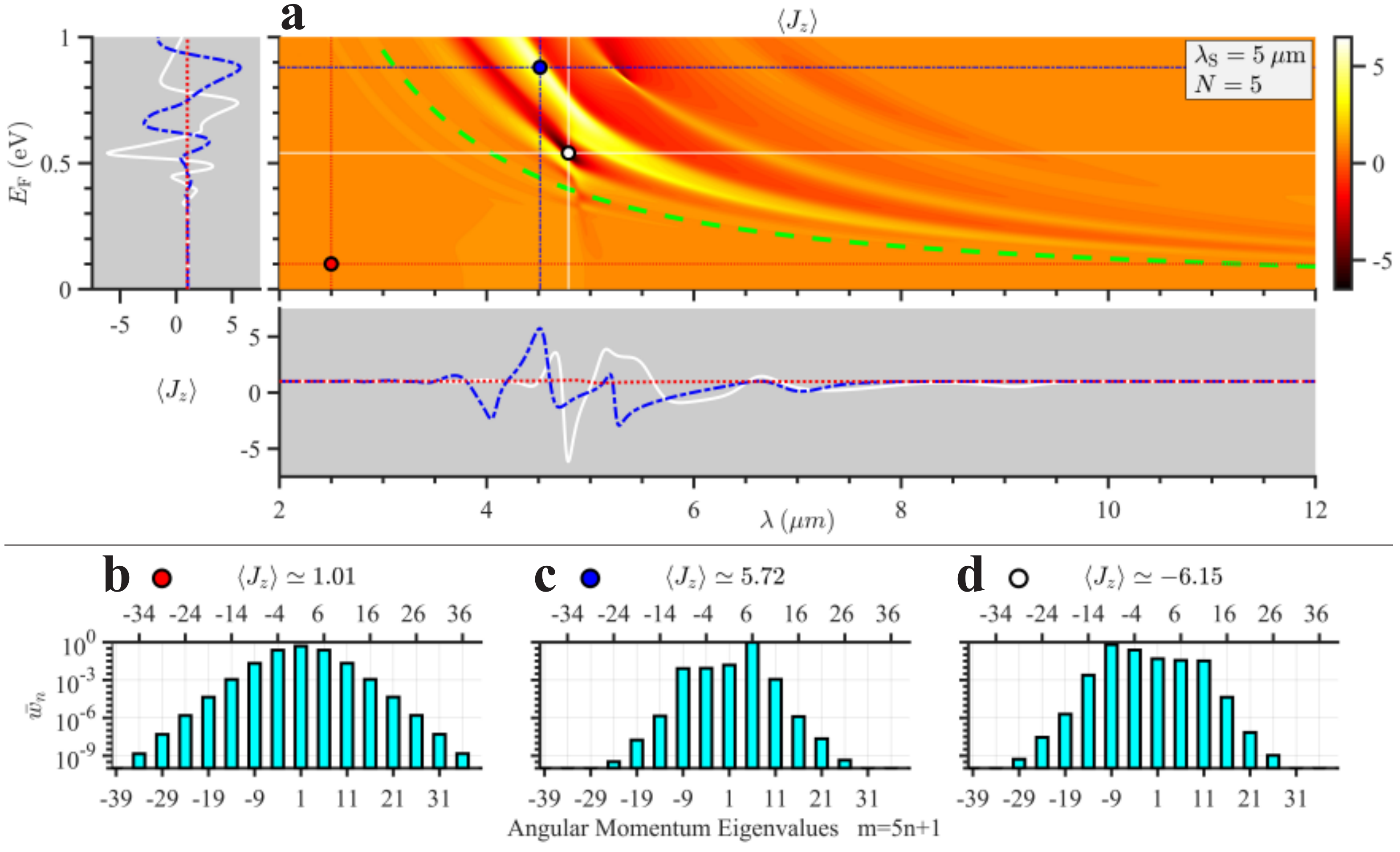}
\caption{{\bf Tunable SAM to OAM conversion.}
  {\bf a} The dependence of the dimensionless average $\left\langle {J_z } \right\rangle$ over the plane $z=0^+$ of the longitudinal angular momentum density (Eq. 6) on the wavelength and the graphene Fermi energy. The green dashed line indicates the GSC threshold as in Fig. 1. Below the GSC threshold $\left\langle {J_z } \right\rangle$ is practically equal to 1 (the impinging angular momentum determined by the SAM only) whereas in the resonant region above the threshold, $\left\langle {J_z } \right\rangle$ spans the range between approximately $-6$ and $6$. The conditions considered in (b--d) and Figs. 4 and 5 are indicated by the coloured disks: (red) a non-resonant case with $\left\langle {J_z } \right\rangle \simeq 1.01$, (blue) and (white) the maximum and minimum average angular momenta $\left\langle {J_z } \right\rangle \simeq 5.72$ and $\left\langle {J_z } \right\rangle \simeq -6.14$, respectively. The left and bottom cross-sections of $\left\langle {J_z } \right\rangle$ correspond to the red, blue and white lines in the main plot. {\bf b}-{\bf d} The distributions $\tilde w_n$ of the scattered angular momentum eigenvalues $m=5n+1$ corresponding to the red, blu and white conditions indicated in {\bf a}. Since $\left\langle {J_z } \right\rangle  = \sum_{n =  - \infty }^{ + \infty } {\left( {nN + 1} \right)\tilde w_n }$, its departure from $1$ (the SAM of the incident plane wave) is produced by the asymmetry of eigenvalues distribution around $m=1$ (n=0).}
\end{figure*}

{\bf Tunable SAM to OAM conversion.} As discussed in the above section, the incident left-polarized plane wave generally excites dipoles at the nanoparticles which are not purely left circularly polarized owing to their mutual interaction and the evanescent coupling with the graphene sheet. Due to their polygonal arrangement with the $C_{N}$ symmetry, these coupled dipoles radiate fields with non-trivial chirality. The densities of the energy, linear and angular momenta of the field in the substrate ($z>0$) can be calculated as \cite{Bliok7}
\begin{eqnarray} \label{densit}
U &=& \frac{1}{4}\left( {{\varepsilon _0}{\varepsilon _2}{{\left| {\bf{E}} \right|}^2} + {\mu _0}{{\left| {\bf{H}} \right|}^2}} \right), \nonumber \\
{\bf{P}} &=& [[{\bf{\hat p}}]] + \frac{1}{2}\nabla  \times [[{\bf{\hat S}}]] \equiv {{\bf{P}}^{\rm o} + {{\bf{P}}^{\rm s}}},  \\
{\bf{J}} &=& [[{\bf{\hat L}}]] +[[{\bf{\hat S}}]], \nonumber
\end{eqnarray}
where ${\bf{\hat p}} = \frac{1}{i}\nabla$, ${\bf{\hat L}} = {\bf{r}} \times {\bf{\hat p}}$ and ${\bf{\hat S}} = {{\bf{e}}_k}\left( {\frac{1}{i}{\varepsilon _{knm}}{{\bf{e}}_n}{{\bf{e}}_m} \cdot } \right)$ are the linear and orbital angular momenta and spin operators, respectively, and we used the notation
\begin{equation}
[[{\bf{\hat O}}]] = \frac{1}{{4\omega }}{\mathop{\rm Re}\nolimits} \left( {{\varepsilon _0}E_j^*{\bf{\hat O}}{E_{j}} + \frac{{{\mu _0}}}{{{\varepsilon _2}}}H_j^*{\bf{\hat O}}{H_j}} \right).
\end{equation}
The linear momentum density ${\bf{P}} = \frac{1}{{2c^2 }}{\mathop{\rm Re}\nolimits} \left( {{\bf{E}} \times {\bf{H}}^* } \right)$ is the well known Belinfante's decomposition \cite{Berry1}, where the orbital contribution ${\bf{P}}^{\rm o} = [[ \hat{\bf p}]]$ is equal to the canonical momentum density, proportional to the local phase gradient. The spin contribution ${\bf{P}}^{\rm s} = \frac{1}{2}\nabla  \times [[{\bf{\hat S}}]]$ corresponds to a transverse spin of evanescent waves \cite{Bliok5}. In order to quantify the SAM to OAM conversion, we consider the quantity (introducing ${\bf{r}}_ \bot   = x{\bf{e}}_x  + y{\bf{e}}_y$)
\begin{equation} \label{Jz}
\left\langle {J_z } \right\rangle  = \omega \varepsilon _2 \frac{\displaystyle {\int {d^2 } {\bf{r}}_ \bot  J_z \left( {{\bf{r}}_ \bot } \right)}}{\displaystyle {\int {d^2 {\bf{r}}_ \bot  } U\left( {{\bf{r}}_ \bot  } \right)}},
\end{equation}
which is a dimensionless average value of the longitudinal angular momentum density $J_z$ on the plane $z=0^+$. From Eq.~(\ref{densit}), it follows
\begin{equation} \label{OAMSAM}
\left\langle {{J_z}} \right\rangle  = \left\langle {{{\bf{r}}_ \bot } \times {\bf{P}}_ \bot ^{\rm o}} \right\rangle  + \left\langle {[[{{\hat S}_z}]]} \right\rangle ,
\end{equation}
which implies that the OAM contribution is entirely due to the orbital part of the linear momentum density \cite{arxiv-19}. Since the eigenvalues of $\hat{S}_z$ are $\pm 1$ and $0$, the SAM contribution is bounded and satisfies the inequality $| { \langle {[[{{\hat S}_z}]]}  \rangle } | \le 1$. Correspondingly, when $|\left\langle {{J_z}} \right\rangle|>1$, the OAM contribution in the left-hand side of Eq.~(\ref{OAMSAM}) is necessarily not vanishing. Below the GSC threshold, $\left\langle{J_z}\right\rangle\simeq~1$ and the scattered field angular momentum is dominated by the SAM retained from the input. However, when the GSC is achieved, $\left\langle {J_z } \right\rangle$ is significantly enhanced $|\left\langle {J_z } \right\rangle|\simeq 6$. The most relevant spectral region for the OAM generation is around the nanoparticle resonance $\lambda_S = 5 \: \mu$m. The SAM to OAM conversion is associated with the OAM generated by the nanostructure-graphene system during the scattering process and the orientation of the induced dipoles. It is now clear that the generated OAM is also tunable as $\left\langle {J_z } \right\rangle$ dramatically depends on 
$E_{\rm F}$, which can be electrically controlled via a gate voltage applied to graphene.

To illustrate such a tunability, we consider three specific configurations as indicated in Fig.~3{\bf a}: (i) a generic non-resonant configuration with $\left\langle {J_z } \right\rangle \simeq 1.01$, and two configurations corresponding to (ii) the maximum $\left\langle {J_z } \right\rangle \simeq 5.72$ and (iii) the minimum $\left\langle {J_z } \right\rangle \simeq -6.14$. The dependence of $\left\langle {J_z } \right\rangle$ on $E_{\rm F}$ is particularly interesting near the resonant wavelengths, which shows that the external gate voltage can continuously tune the scattered angular momentum. The weights of the total angular momentum eigenstates generated by the polygonal array depend on the dipole orientation ${\bf p}^{(0)}$. Owing to the $C_N$ symmetry, only eigenstates with eigenvalues $m=nN+1$ are excited, where $n$ is an integer (see Supplementary information). The average angular momentum can be expressed as a weighted average of the eigenvalues
\begin{equation}
\left\langle {J_z } \right\rangle  = \sum\limits_{n =  - \infty }^{ + \infty } {\left( {nN + 1} \right)\tilde w_n },
\end{equation}
where $\tilde w_n$ is the normalized distribution of angular momentum eigenvalues. As we can see from the distributions of the eigenvalues $m=5n+1$, $\tilde w_n$ goes rapidly to zero as $|n|$ increases. In the nonresonant configuration (Fig.~3{\bf b}), the distribution is nearly symmetrical with respect to $m=1$ ($n=0$) and the average is very close to $1$. In the resonant configuration (Fig.~3{\bf c,d}), the distributions are not symmetric around $m=1$ and for $\langle J_z \rangle\simeq 6$ is dominated by the eigenvalue $m=6$, while for $\langle J_z \rangle\simeq -6$, the most important are the eigenvalues $m=-9$ and $m=-4$.
\begin{figure*} \label{Fig4}
\center
\includegraphics[width=1\textwidth]{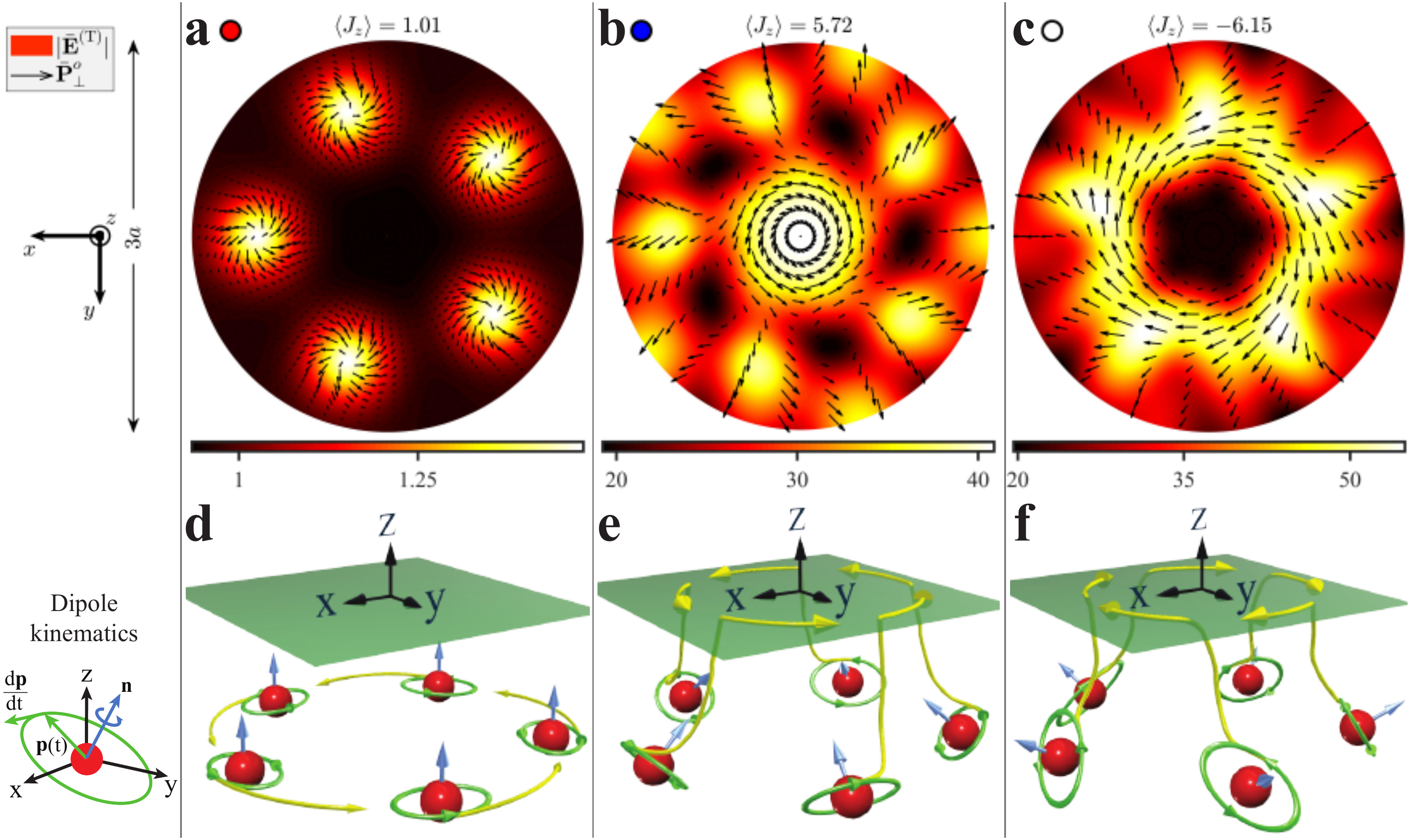}
\caption{ {\bf Excitation of GPPs carrying OAM}.
  {\bf a}-{\bf c} The spatial distributions around the polygon ($r_\bot < 3a$) of the normalized electric field magnitude $|{{{\bf{\bar E}}}^{\left( \rm T \right)}}| =| {\bf{E}}/{E_0}|$ and transverse orbital linear momentum density $\bar {\bf{P}} _ \bot ^{\rm o} = {\bf P}_\bot / P_0$ in the plane $z=0^+$ under the conditions indicated in Fig. 3. In the non-resonant conditions {\bf a}, the electric field does not show an overall spatial chirality, $\bar {\bf{P}} _ \bot ^{\rm o}$ does not circulate around the $z$-axis, and there is no net angular momentum exchange upon scattering. In the resonant conditions {\bf b} and {\bf c}, the electric fields patterns have a marked spatial two-dimensional chiral trait. The vector fields $\bar {\bf{P}} _ \bot ^{\rm o}$ swirl around the $z$-axis and they produce OAM contributions $\left\langle {{{\bf{r}}_ \bot } \times {\bf{P}}_ \bot ^{\rm o}} \right\rangle$  which agree with the signs of the average angular momenta $\langle J_z \rangle$. {\bf d}-{\bf f} Pictorial representation of the dipole polarization ellipses and corresponding GPP excitation on the graphene sheet (green plane) under the conditions corresponding to {\bf a}-{\bf c}. In the non-resonant configuration {\bf d}, the polarization ellipses are parallel to the graphene sheet and the dipoles do not excite GPPs. In the resonant configurations {\bf e} and {\bf f}, the dipole spins precess with inward and outward orientation towards the polygon centre and excite GPPs traveling in the anticlockwise or clockwise in azimuthal direction. Their interference results in the GPP field on the graphene plane positively and negatively rotating around the $z$-axis, which generate the OAM carried by the scattered field.}
\end{figure*}
\begin{figure*} \label{Fig5}
\center
\includegraphics[width=1\textwidth]{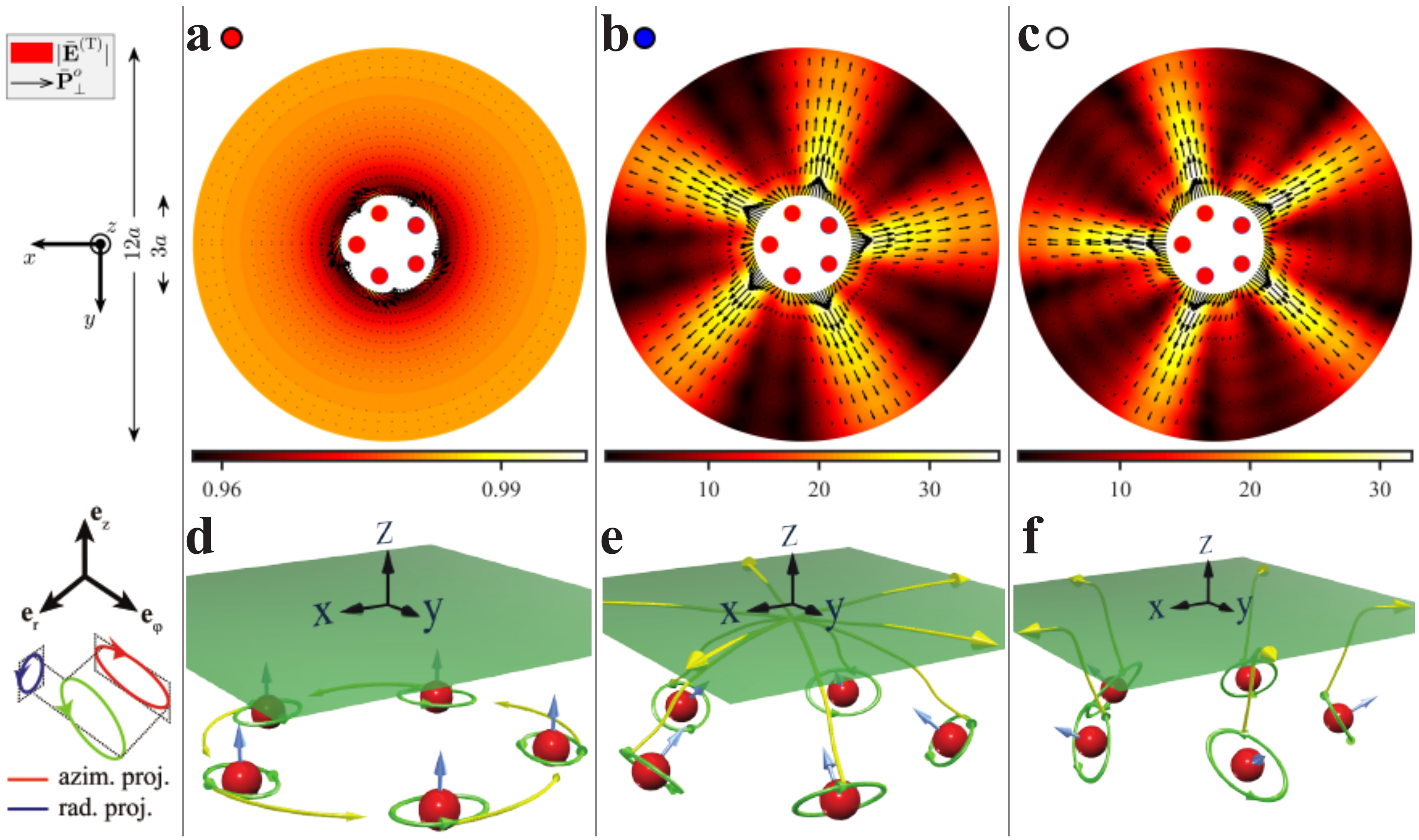}
\caption{{\bf Radial excitation of GPPs}. {\bf a}-{\bf c}
The spatial distributions around the polygon ($3a < r_\bot < 12a$) of the normalized electric field magnitude $|{{{\bf{\bar E}}}^{\left( \rm T \right)}}| =| {\bf{E}}/{E_0}|$ and transverse orbital linear momentum density $\bar {\bf{P}} _ \bot ^{\rm o} = {\bf P}_\bot / P_0$ in the plane $z=0^+$
  under the conditions indicated in Fig. 3 (cf. Fig. 4). In the non-resonant conditions {\bf a}, GPPs excitation is negligible. In the resonant conditions {\bf b} and {\bf c}, each nanoparticle is responsible for the excitation of a single GPP propagating along the corresponding polygon apothem. These GPPs are excited by the radial projections of the dipolar moments. In the resonant conditions {\bf e} and {\bf f}, the radial projections of the dipole moments negatively and positively rotate in the radial plane, respectively. Accordingly in {\bf e} GGPs are launched from the nanoparticles toward the polygon center whereas in {\bf f} the situation is reversed.}
\end{figure*}

{\bf Excitation of GPPs with defined rotation direction.} It is interesting to note that during the observed SAM-to-OAM conversion, the scattered angular momentum can have opposite sign with respect the incoming SAM: {\it the scattering can flip chirality of light}. In order to examine this effect in more detail, we evaluate the electric field and the transverse orbital linear momentum density in the plane $z=0^+$, $|{{{\bf{\bar E}}}^{\left( \rm T \right)}}| =| {\bf{E}}\left( {{{\bf{r}}_ \bot }} \right)/{E_0}|$ and $\bar {\bf{P}} _ \bot ^{\rm o} = \left[ {P_x^{\rm o}\left( {{{\bf{r}}_ \bot }} \right){{\bf{e}}_x} + P_y^{\rm o}\left( {{{\bf{r}}_ \bot }} \right){{\bf{e}}_y}} \right]/ P_0$, normalized by the impinging field magnitude $|E_0|$ and linear momentum density $P_0 = {\frac{1}{{2{c^2}}}\sqrt {\frac{{{\varepsilon _0}{\varepsilon _1}}}{{{\mu _0}}}} {{\left| {{E_0}} \right|}^2}}$, respectively. The spatial shapes of these real quantities are invariant under discrete rotations (by angles $2 \pi r/N$) around the $z$ axis. In the non-resonant case (Fig. 4{\bf a}), the scattered electric field nearly coincides with the incoming field, with the exception of five lobes, spatially overlapping the position of the nanoparticles. The orbital linear momentum $\bar {\bf{P}} _ \bot ^{\rm o}$ spreads out of each lobe and is negligible everywhere else, therefore, yielding a very small OAM contribution $\left\langle {{{\bf{r}}_ \bot } \times {\bf{P}}_ \bot ^{\rm o}} \right\rangle$ in Eq.~(\ref{OAMSAM}). Accordingly, $\langle J_z \rangle\simeq 1$ and the angular momentum exchange upon scattering is absent. The situation is completely different in the resonant configurations (Fig. 4{\bf b},{\bf c}). The electric field magnitude is an order of magnitude larger than the input (due to the plasmonic field enhancement),
and the pattern has the shape of a ratchet wheel with a marked two-dimensional chiral trait. The vector fields $\bar {\bf{P}} _ \bot ^{\rm o}$ swirl around the $z$-axis and their positive and negative swirling directions produce positive and negative OAM contributions ($\left\langle {{{\bf{r}}_ \bot } \times {\bf{P}}_ \bot ^{\rm o}} \right\rangle$ in Eq.~(\ref{OAMSAM})). This is in agreement with the sign of the average angular momentum in each case. Because the orbital linear momentum density ${\bf P}^{\rm o}$ produces a mechanical force on an absorbing particle, these results suggest a novel way of manipulating the motion and rotation of nanoparticles through electric gating of a graphene substrate.

The SAM to OAM conversion mechanism has a simple and clear physical interpretation in terms of unidirectional excitation of GPPs with defined rotation direction. The transmitted angular momentum depends on the dipole moments of the nanoparticles forming the polygon, which are all rotated doppelgangers of ${\bf p}^{(0)}$ with a geometric phase factor (Eq.~(\ref{invar})). The time-dependent dipole moment ${\bf p} (t)= {\rm Re} \left[{\bf p}^{(0)} e^{-i \omega t} \right]$ sweeps a polarization ellipse whose spatial orientation and rotation direction are given by the time-independent angular normal vector
\begin{equation}
{\bf{n}} = {\bf{p}} \times \frac{{d{\bf{p}}}}{{dt}} = \frac{{i\omega }}{2}{{\bf{p}}^{\left( 0 \right)}} \times {{\bf{p}}^{\left( 0 \right)*}},
\end{equation}
as shown in Fig. 4 (Dipole kinematics). In the strong coupling regime, we obtain (see Supplementary Information)
\begin{equation} \label{Jzz}
\left\langle {{J_z}} \right\rangle  = 1 - \frac{\gamma }{\omega }{n_x}
\end{equation}
with $\gamma>0$. Equation~(\ref{Jzz}) states that the orientation of the polarization ellipse of ${\bf p}^{(0)}$ determines the scattered angular momentum, which increases (decreases) the incident light SAM if the polarization ellipse is negatively (positively) oriented with respect the $x$-axis. The polarization ellipse of the $q$-th dipole coincides with the $0$-th ellipse rotated by an angle $2 \pi q /N$. If $n_x<0$, all the normal vectors point towards the $z$-axis and the opposite for $n_x>0$ (Eq.(\ref{invar})). In the non-resonant configuration (Fig. 4{\bf d}), the polarization ellipses are parallel to the graphene sheet as the dipole ${\bf p}^{(0)}$ is nearly left polarized with $n_x \simeq 0$ and unitary angular momentum. The dipoles rotating parallel to the graphene sheet provide negligible excitation of GPPs. This changes in the resonant configurations for $n_x <0$ and $\langle J_z \rangle >0$ (Fig. 4{\bf e}) and $n_x >0$ and $\langle J_z \rangle <0$ (Fig. 4{\bf f}). Equation~(\ref{Jzz}) supports the interpretation of the SAM to OAM conversion in terms of excitation of GPPs \cite{Rodri1}. In the configuration in Fig.~4{\bf e}, each dipole rotates with the normal vector $\bf n$ pointing towards the $z$-axis and excites respective GPP, which travels in the positive azimuthal direction (yellow arrows in Fig. 4). The interference of all these azimuthally traveling GPPs produces a scattered electromagnetic field rotating around the $z$-axis in the positive direction, which is responsible for the positive OAM of the scattered field.  The reversed situation occurs in the configuration of Fig.~4{\bf f}, where the normal vectors $\bf n$ point in opposite direction of the $z$-axis, and the excited GPP travels in the opposite direction to the previous case.

{\bf Discrete radial emission of GPPs.} The nanophotonic mechanism supporting the SAM to OAM conversion is also responsible for another effect of shaping GPPs beams
excited by the nanoparticles outside the polygon central region (although it is not connected to SAM-OAM exchange). In the non-resonant configuration (Fig. 5{\bf a}), the scattered electric field is nearly equal to the incident one and the transverse orbital linear momentum density is vanishing. In the resonant case (Fig. 5{\bf b},{\bf c}), the electric field is enhanced within specific circular sectors over which the orbital linear momentum density is almost radial. These circular sectors are located around the polygon 
apothem lines, along the direction from the nanospheres to the polygon center (Fig.~5{\bf b}) and from the center to nanoparticles (Fig.~5{\bf c}). The circular sectors, as the whole region $r_\bot > 3a$, provide a vanishing OAM contribution since ${{{\bf{r}}_ \bot } \times {\bf{P}}_ \bot ^{\rm o}} \simeq 0$ everywhere, so that the effect is independent on the above described angular momentum exchange. A detailed analysis of the transverse asymptotic field behavior (see Supplementary Information) indicates that each circular section hosts the GPP excited by an individual nanosphere and traveling along the corresponding polygon apothem. Such discrete radial emission of GPPs can be understood in terms of unidirectional excitation and interference of graphene plasmon modes produced by the rotating dipoles. Since ${{\bf{p}}^{\left( 0 \right)}} = p_r^{\left( 0 \right)}{{\bf{e}}_r} + p_\varphi ^{\left( 0 \right)}{{\bf{e}}_\varphi } + p_z^{\left( 0 \right)}{{\bf{e}}_z}$, where ${\bf{e}}_r$, ${\bf{e}}_\varphi$ and ${\bf{e}}_z $ are the cylindrical coordinates unit vectors, each polarization ellipse has both azimuthal and radial projections swept by the rotating dipole of complex amplitudes $p_\varphi ^{\left( 0 \right)}{{\bf{e}}_\varphi } + p_z^{\left( 0 \right)}{{\bf{e}}_z}$ and $p_r^{\left( 0 \right)}{{\bf{e}}_r} + p_z^{\left( 0 \right)}{{\bf{e}}_z}$, respectively (Fig. 5) The azimuthal projections of the ellipses are responsible for the excitation of the azimuthally propagating GPPs. These GPPs have finite transverse width in the radial direction, and the azimuthal projections of the ellipses play no role in the region $r_\bot > 3a$. On the other hand, the radial projections of the rotating dipoles are able to excite GPPs along the polygon apothems. While in the non-resonant configuration (Fig.~5{\bf d}) GPP excitation is negligible, in the resonant case (Fig.~5{\bf e}), the polarization ellipses are inward inclined and this implies that the radial projections of the rotating dipoles launch GPPs traveling along the direction from each nanoparticle toward the polygon center. Conversely, in the case of Fig.~5{\bf f}, the polarization ellipses are outward inclined and accordingly the radial projections of the rotating dipoles launch GPPs propagating along the direction from the polygon center toward each nanoparticle.

We also stress that such GPPs radial emission is strongly affected by the Fermi level in terms of efficiency (in Fig.~5{\bf a} there is no emission) and directionality (in Figs.~5{\bf b} and 5{\bf c}, GPPs travel along opposite directions of the polygon apothem lines). This implies that the beaming of the radial GPPs can be tuned by gated electrical signals, in the same way as the OAM of the scattered light described above.

{\bf Conclusions}

We have described electrically tunable SAM to OAM conversion at the nanoscale using a plasmonic nanoparticle arrangement with broken rotational invariance placed on a graphene substrate. A polygonal array of ITO nanospheres deposited on top of a graphene sheet generates scattered light carrying net OAM under illumination by a plane wave carrying only SAM. The effect originates from the rotated induced dipoles of the nanoparticles due to the strong coupling between LSPs of the nanoparticles and graphene plasmons and, therefore, resonant and strongly depends on the graphene Fermi energy. This opens up a possibility to continuously tune the generated OAM via electric biasing of the graphene sheet. The novel nanophotonic mechanism supporting this effect results from the interplay of two physical mechanisms. First, the hybridization of the LSPs of the nanoparticle and GPPs entails excitation of the nanoparticle dipoles whose rotation plane is not parallel to the graphene sheet. Second, the inclined rotating dipoles excite GPPs traveling in the azimuthal direction around the polygon centre on the graphene so that their interference produces a field distribution with net OAM. The same spin-orbit interaction results also in a controlled emission of graphene plasmons radially propagating along the polygon apothem lines. The amplitude and propagation direction of these strongly confined plasmonic beams can also be controlled by electrically doping the graphene layer.

The proposed hybrid graphene-plasmonic nanostructure reveals a rich new framework for studying many new physical phenomena driven by the strong field localization, and unexpected couplings with non-trivial plasmonic resonances and electromagnetic field topologies. The described effect may trigger many possibilities for the precise control of information and beam propagation at the nanoscale in 2D materials and may also be extended to nonlinear and quantum regimes. The developed approach opens up new avenues for manipulating OAM of light at the nanoscale and the development of integrated devices for optical manipulation, information processing, communication, and classical and quantum computing at the nanoscale.

{\bf Acknowledgments}

A.C and C.C. acknowledge PRIN 2017 PELM (grant number 20177PSCKT). A.M. acknowledges support from the Rita Levi Montalcini Fellowship (grant number PGR15PCCQ5) funded by the Italian Ministry of Education, Universities and Research (MIUR). A.Z. acknowledges support from the EPSRC (UK) and the ERC iCOMM project (789340).

{\bf Author Contributions}

A.C. conceived the idea and worked out the theory. All the authors discussed the results and wrote the paper.

{\bf Competing financial interests}

The authors declare no competing financial interest.

\end{document}


\title{Electric control of spin orbit coupling in graphene-based nanostructures with broken rotational symmetry: Supplementary Information}
\author{A. Ciattoni$^1$}
\email{alessandro.ciattoni@aquila.infn.it}
\author{C. Conti$^{2,3}$}
\author{A. V. Zayats$^4$}
\author{A. Marini$^5$}
\email{andrea.marini@aquila.infn.it}
\affiliation{$^1$CNR-SPIN, c/o Dip.to di Scienze Fisiche e Chimiche, Via Vetoio, 67100 Coppito (L'Aquila), Italy}
\affiliation{$^2$CNR-ISC, Via dei Taurini 19, 00185, Rome, Italy}
\affiliation{$^3$Department of Physics, University Sapienza, Piazzale Aldo Moro 5, 00185, Rome, Italy}
\affiliation{$^4$Department of Physics, King’s College London, Strand, London, WC2R 2LS, United Kingdom }
\affiliation{$^5$Department of Physical and Chemical Sciences, University of L'Aquila, Via Vetoio, 67100 L'Aquila, Italy}
\date{\today}
\begin{abstract}

\end{abstract}

\maketitle

\section{Dipoles coupling and excitation}
Consider a graphene sheet lying on the plane $z=0$ between the half spaces $z<0$ and $z>0$ filled by two transparent dielectrics of permittivities $\epsilon_1$ and $\epsilon_2$ (and wavenumbers $k_j  = k_0 \sqrt {\varepsilon _j }$, $k_0 = \omega/c$), respectively (see Fig.1). Dielectric 1 hosts $N$ identical spheres, with radius $R$ and dielectric permittivity $\varepsilon_{\rm S} (\omega)$, lying on top of the graphene sheet and with centers arranged at the vertices of a regular polygon of radius $a$, whose positions are
\begin{equation} \label{rq}
{\bf{r}}^{\left( q \right)}  = a\left[ {\cos \left( {\frac{{2\pi }}{N}q} \right){\bf{e}}_x  + \sin \left( {\frac{{2\pi }}{N}q} \right){\bf{e}}_y } \right] - R{\bf{e}}_z,
\end{equation}
where $q=0,\ldots,N-1$ and ${\bf{e}}_x$, ${\bf{e}}_y$, ${\bf{e}}_z$ are the cartesian unit vectors. Consider now a monochromatic ($e^{-i \omega t}$) plane wave ${\bf{E}}_{0 \bot } e^{ik_1 z}$ (where ${\bf{e}}_z \cdot {\bf{E}}_{0 \bot }  = 0 $) launched from dielectric 1 and normally impinging upon the spheres-graphene system. In order to investigate such scattering process, we focus on the relevant situation where the spheres' radius is much smaller than the vacuum wavelength $R \ll \lambda  = {2\pi c}/{\omega }$. In this situation one can resort to the quasi-electrostatic approximation where the $q$-th sphere is modelled by a point dipole located at ${\bf{r}}^{\left( q \right)}$ and with dipole moment
\begin{equation} \label{dipoleMoment}
{\bf{p}}^{\left( q \right)}  = \alpha {\bf{E}}_q^{\left( {\rm Ext} \right)},
\end{equation}
where ${\bf{E}}_q^{\left( {\rm Ext} \right)}$ is the local field experienced by the dipole without self-field and
\begin{equation}
\alpha  = 4\pi \varepsilon _0 \varepsilon _1 R^3 \left( {\frac{{\varepsilon _{\rm S}  - \varepsilon _1 }}{{\varepsilon _{\rm S}  + 2\varepsilon _1 }}} \right)
\end{equation}
is the well-known polarizability of the sphere. The field radiated by the $q$-th dipole is
\begin{equation} \label{dipoleField}
{\bf{E}}^{\left( q \right)} \left( {\bf{r}} \right) = \left( {k_1^2  + \nabla \nabla  \cdot } \right)\left( {\frac{1}{{4\pi \varepsilon _0 \varepsilon_1 }}\frac{{e^{ik_1 \left| {{\bf{r}} - {\bf{r}}^{\left( q \right)} } \right|} }}{{\left| {{\bf{r}} - {\bf{r}}^{\left( q \right)} } \right|}}{\bf{p}}^{\left( q \right)} } \right),
\end{equation}
so that the overall field in the half space $z<0$ impinging onto the graphene layer is
\begin{equation} \label{IncidentField}
{\bf{E}}^{\left(  \rm I \right)} \left( {\bf{r}} \right) = {\bf{E}}_{0 \bot } e^{ik_1 z}  + \sum\limits_{q = 0}^{N - 1} {{\bf{E}}^{\left( q \right)} \left( {\bf{r}} \right)},
\end{equation}
whereas the graphene sheet and the dielectric discontinuity at $z=0$ produce the reflected and transmitted fields, ${\bf{E}}^{\left( \rm  R \right)} \left( {\bf{r}} \right)$ and ${\bf{E}}^{\left(  \rm T \right)} \left( {\bf{r}} \right)$, in the half spaces $z<0$ and $z>0$ respectively (see below). The overall field at the $q$-th dipole without self-field is
\begin{equation}
{\bf{E}}_q^{\left( {\rm Ext} \right)}  = \left[ {{\bf{E}}^{\left( \rm I \right)} \left( {\bf{r}} \right) + {\bf{E}}^{\left( \rm R \right)} \left( {\bf{r}} \right) - {\bf{E}}^{\left( q \right)} \left( {\bf{r}} \right)} \right]_{{\bf{r}} = {\bf{r}}^{\left( q \right)} }
\end{equation}
so that Eq.(\ref{dipoleMoment}) becomes
\begin{equation} \label{consistency}
\frac{1}{\alpha }{\bf{p}}^{\left( q \right)}  = {\bf{E}}_{0 \bot } e^{ - ik_1 R}  + \sum\limits_{\scriptstyle s = 0 \hfill \atop
  \scriptstyle s \ne q \hfill}^{N - 1} {{\bf{E}}^{\left( s \right)} \left( {{\bf{r}}^{\left( q \right)} } \right)}  + {\bf{E}}^{\left( \rm R \right)} \left( {{\bf{r}}^{\left( q \right)} } \right),
\end{equation}
i.e. a dipole is immersed in the field of the incident plane wave, the fields of the other $N-1$ dipoles and the field reflected by graphene and dielectric discontinuity.

\begin{figure*} \label{Fig1}
\center
\includegraphics[width=0.9\textwidth]{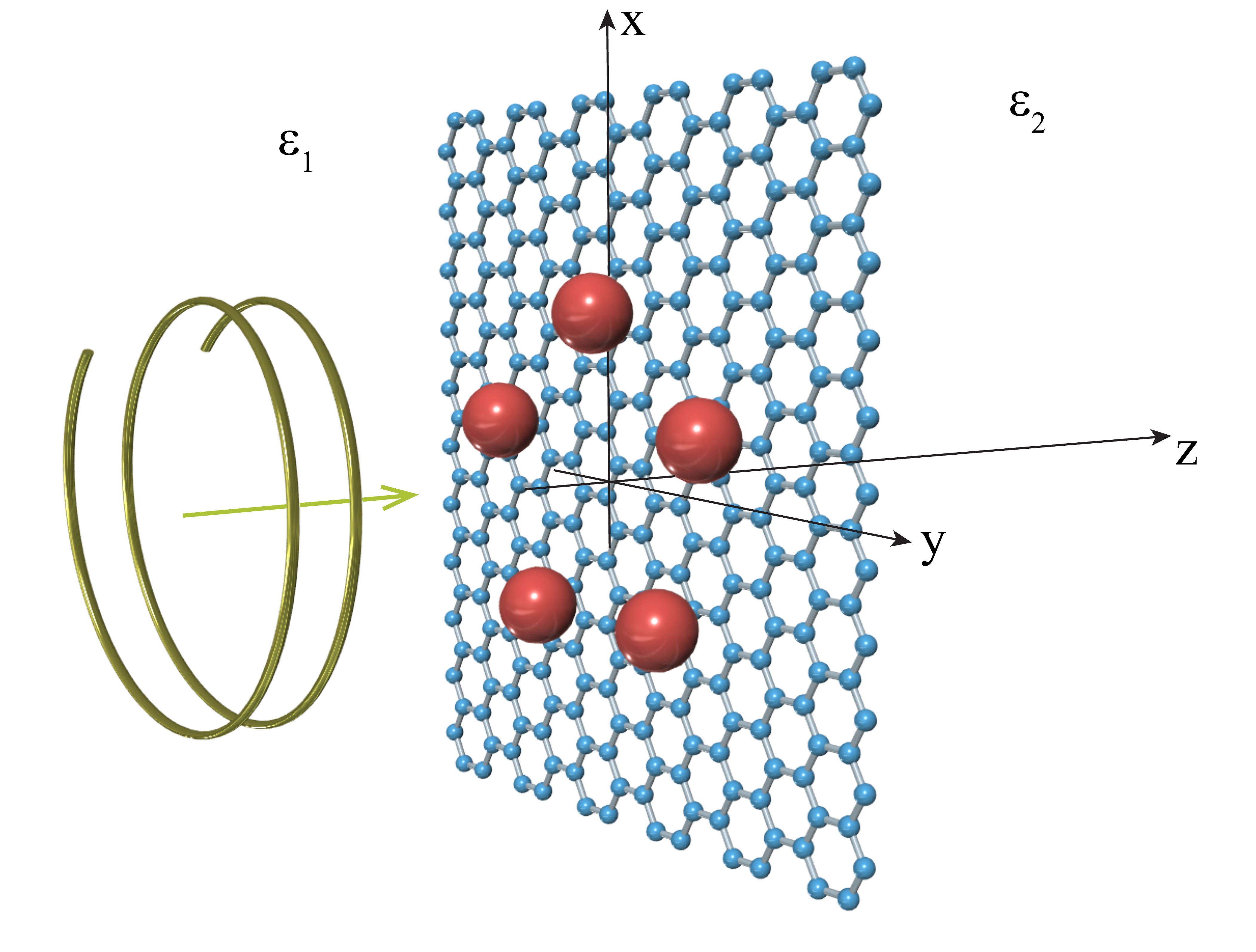}
\caption{Geometry of the sphere-graphene system.}
\end{figure*}

In order to obtain the equations governing the dipole excitation, we note that from Eq.(\ref{dipoleField}) we get
\begin{equation} \label{dipoleField1}
{\bf{E}}^{\left( s \right)} \left( {{\bf{r}}^{\left( q \right)} } \right) = \frac{{k_1^3 }}{{4\pi \varepsilon _0 \varepsilon _1 }}\left\{ {\frac{{e^{iu} }}{u}\left[ {\left( {\frac{{ - 1 + iu + u^2 }}{{u^2 }}} \right) + \left( {\frac{{3 - 3iu - u^2 }}{{u^2 }}} \right)\frac{{{\bf{uu}}^T }}{{u^2 }}} \right]} \right\}_{{\bf{u}} = k_1 \left( {{\bf{r}}_ \bot ^{\left( s \right)}  - {\bf{r}}_ \bot ^{\left( q \right)} } \right)} {\bf{p}}^{\left( s \right)}
\end{equation}
where we hereafter use the dyadic notation $\left( {{\bf{ab}}^T } \right){\bf{v}} = {\bf{a}}\left( {{\bf{b}} \cdot {\bf{v}}} \right)$. On the other hand, to evaluate the reflected and the transmitted fields it is convenient to resort to the angular spectrum formalism. We hereafter adopt the notation ${\bf F}_\bot = {\bf F} - F_z {\bf e}_z$ to label the transverse part of a vector ${\bf F}$ and we set
\begin{equation}
k_{jz} \left( {{\bf{k}}_ \bot  } \right) = \sqrt {k_j^2  - k_ \bot ^2 }  = \left\{ \begin{array}{l}
 \sqrt {k_j^2  - k_ \bot ^2 }, \quad \quad k_ \bot   < k_j  \\
 i\sqrt {k_ \bot ^2  - k_j^2 }, \quad \quad k_ \bot   > k_j  \\
 \end{array} \right.
\end{equation}
for the longitudinal components of the wave vectors in the dielectrics $j=1,2$. Representing the incident field as
\begin{equation}
{\bf{E}}^{\left( \rm I \right)} ({\bf r}) = \int {d^2 {\bf{k}}_ \bot  e^{i{\bf{k}}_ \bot   \cdot {\bf{r}}_ \bot  } } e^{ik_{1z} z} \left( {1 - \frac{{{\bf{e}}_z {\bf{k}}_ \bot ^T  }}{{k_{1z} }}} \right){\bf{\tilde E}}_ \bot ^{\left( \rm I \right)} ({\bf k}_\perp),
\end{equation}
the reflected and transmitted fields are
\begin{eqnarray} \label{RefTraFields}
 {\bf{E}}^{\left( \rm R \right)} ({\bf r}) &=& \int {d^2 {\bf{k}}_ \bot  } e^{i{\bf{k}}_ \bot   \cdot {\bf{r}}_ \bot  } e^{ - ik_{1z} z} \left( {1 + \frac{{{\bf e }_z {\bf{k}}_ \bot ^T  }}{{k_{1z} }}} \right)\left[ {r_{\rm TE} \left( {1 - \frac{{{\bf{k}}_ \bot  {\bf{k}}_ \bot ^T  }}{{k_ \bot ^2 }}} \right) + r_{\rm TM} \left( {\frac{{{\bf{k}}_ \bot  {\bf{k}}_ \bot ^T  }}{{k_ \bot ^2 }}} \right)} \right]{\bf{\tilde E}}_ \bot ^{\left( \rm I \right)} ({\bf k}_\perp), \nonumber  \\
 {\bf{E}}^{\left( \rm T \right)} ({\bf r}) &=& \int {d^2 {\bf{k}}_ \bot  e^{i{\bf{k}}_ \bot   \cdot {\bf{r}}_ \bot  } } e^{ik_{2z} z} \left( {1 - \frac{{{\bf e}_z {\bf{k}}_ \bot ^T  }}{{k_{2z} }}} \right)\left[ {t_{\rm TE} \left( {1 - \frac{{{\bf{k}}_ \bot  {\bf{k}}_ \bot ^T  }}{{k_ \bot ^2 }}} \right) + t_{\rm TM} \left( {\frac{{{\bf{k}}_ \bot  {\bf{k}}_ \bot ^T  }}{{k_ \bot ^2 }}} \right)} \right]{\bf{\tilde E}}_ \bot ^{\left( \rm I \right)} ({\bf k}_\perp),
\end{eqnarray}
where the transverse electric ($\rm TE$) and transverse magnetic ($\rm TM$) reflectances ($r$) and transmittances ($t$) are
\begin{eqnarray} \label{rt}
r_{\rm TE} \left( {k_ \bot  } \right) &=& \frac{{k_{1z}  - k_{2z}  - k_0 Z_0 \sigma }}{{k_{1z}  + k_{2z}  + k_0 Z_0 \sigma }}, \nonumber \\
 r_{\rm TM} \left( {k_ \bot  } \right) &=&  \frac{\displaystyle {\frac{{\varepsilon _1 }}{{k_{1z} }} - \frac{{\varepsilon _2 }}{{k_{2z} }} - \frac{{Z_0 \sigma }}{{k_0 }}}}{\displaystyle  {\frac{{\varepsilon _1 }}{{k_{1z} }} + \frac{{\varepsilon _2 }}{{k_{2z} }} + \frac{{Z_0 \sigma }}{{k_0 }}}}, \nonumber \\
\displaystyle t_{\rm TE} \left( {k_ \bot  } \right) &=& \frac{{2k_{1z} }}{{k_{1z}  + k_{2z}  + k_0 Z_0 \sigma }}, \nonumber \\
\displaystyle t_{\rm TM} \left( {k_ \bot  } \right) &=& \frac{\displaystyle  {2\frac{{\varepsilon _1 }}{{k_{1z} }}}}{\displaystyle {\frac{{\varepsilon _1 }}{{k_{1z} }} + \frac{{\varepsilon _2 }}{{k_{2z} }} + \frac{{Z_0 \sigma }}{{k_0 }}}},
\end{eqnarray}
where $\sigma$ is the graphene surface conductivity and  $Z_0  = \sqrt {\mu _0 /\varepsilon _0 }$ is the vacuum impedance. Using the Weyl representation of the spherical wave

\begin{equation}
\frac{{e^{ik_1 r} }}{r} = \frac{i}{{2\pi }}\int {d^2 {\bf{k}}_ \bot  e^{i{\bf{k}}_ \bot   \cdot {\bf{r}}_ \bot  } } \frac{{e^{ik_{1z} \left| z \right|} }}{{k_{1z} }},
\end{equation}
from Eq.(\ref{dipoleField}) and Eq.(\ref{IncidentField}), we readily get
\begin{equation} \label{incidentspectrum}
{\bf{\tilde E}}_ \bot ^{\left( \rm I \right)} \left( {{\bf{k}}_ \bot  } \right) = {\bf{E}}_{0 \bot } \delta \left( {{\bf{k}}_ \bot  } \right) + \frac{i}{{8\pi ^2 \varepsilon _0 \varepsilon _1 }}\frac{{e^{ik_{1z} R} }}{{k_{1z} }}\sum\limits_{s = 0}^{N - 1} {e^{ - i{\bf{k}}_ \bot   \cdot {\bf{r}}_ \bot ^{\left( s \right)} } \left( {k_1^2 I_ \bot   - {\bf{k}}_ \bot  {\bf{k}}_ \bot ^T  - k_{1z} {\bf{k}}_ \bot  {\bf{e}}_z^T } \right){\bf{p}}^{\left( s \right)} }
\end{equation}
which inserted into the first of Eq.(\ref{RefTraFields}) yields the reflected field. Using Eq.(\ref{dipoleField1}) and the first of Eq.(\ref{RefTraFields}), Eq.(\ref{consistency}) turns into
\begin{equation} \label{dipoleexcitation}
{\beta {\bf{p}}^{\left( q \right)}  - \sum\limits_{s = 0}^{N - 1} {H^{\left( {q,s} \right)} } {\bf{p}}^{\left( s \right)}  = \left[ {e^{ - ik_1 R}+ e^{ik_1 R} r_{\rm TE} \left( 0 \right)   } \right]\left( {4\pi \varepsilon _0 \varepsilon _1 R^3 } \right){\bf{E}}_{0 \bot } }
\end{equation}
where
\begin{eqnarray} \label{couplings}
\beta  &=& \frac{{4\pi \varepsilon _0 \varepsilon _1 R^3 }}{\alpha } \equiv \frac{{\varepsilon _{\rm S}  + 2\varepsilon _1 }}{{\varepsilon _{\rm S}  - \varepsilon _1 }}, \nonumber \\
 H^{\left( {q,s} \right)}  &=& \left[ {\left( {1 - \delta _{q,s} } \right)G^{\left( {\rm Int} \right)} \left( {{\bf{r}}_ \bot  } \right) + G_{\rm TE}^{\left( \rm Ref \right)}\left( {{\bf{r}}_ \bot  } \right) + G_{\rm TM}^{\left( \rm Ref \right)}\left( {{\bf{r}}_ \bot  } \right)} \right]_{{\bf{r}}_ \bot   = {\bf{r}}_ \bot ^{\left( q \right)}  - {\bf{r}}_ \bot ^{\left( s \right)} }, \nonumber \\
 G^{\left( {\rm Int} \right)} \left( {{\bf{r}}_ \bot  } \right) &=& \left( {k_1 R} \right)^3 \frac{{e^{ik_1 r_ \bot  } }}{{k_1 r_ \bot  }}\left[ {\left( {1 + \frac{{ik_1 r_ \bot   - 1}}{{k_1^2 r_ \bot ^2 }}} \right) + \left( {\frac{{3 - 3ik_1 r_ \bot   - k_1^2 r_ \bot ^2 }}{{k_1^2 r_ \bot ^2 }}} \right)\frac{{{\bf{r}}_ \bot  {\bf{r}}_ \bot ^T }}{{r_ \bot ^2 }}} \right], \nonumber \\
 G_{\rm TE}^{\left( \rm Ref \right)}\left( {{\bf{r}}_ \bot  } \right) &=& \frac{{iR^3 }}{{2\pi  }}\int {d^2 {\bf{k}}_ \bot  e^{i{\bf{k}}_ \bot   \cdot {\bf{r}}_ \bot  } } e^{i k_{1z} 2R} \frac{{k_1^2 }}{{k_{1z} }}r_{\rm TE} \left( {I_ \bot   - \frac{{{\bf{k}}_ \bot  {\bf{k}}_ \bot ^T }}{{k_ \bot ^2 }}} \right), \nonumber \\
 G_{\rm TM}^{\left( \rm Ref \right)}\left( {{\bf{r}}_ \bot  } \right) &=& \frac{{iR^3 }}{{2\pi  }}\int {d^2 {\bf{k}}_ \bot  e^{i{\bf{k}}_ \bot   \cdot {\bf{r}}_ \bot  } } e^{i k_{1z} 2R} \frac{{k_ \bot ^2 }}{{k_{1z} }}r_{\rm TM} \left( {\frac{{k_{1z} }}{{k_ \bot ^2 }}{\bf{k}}_ \bot   + {\bf{ e}}_z } \right)\left( {\frac{{k_{1z} }}{{k_ \bot ^2 }}{\bf{k}}_ \bot   - {\bf{ e}}_z } \right)^T,
\end{eqnarray}

Equation (\ref{dipoleexcitation}) is one of the main result of the present Supplementary Information since it is the basic equation describing the excitation process of the dipoles representing the spheres. As a matter of fact, for $q=0,\ldots,N-1$, Eq. (\ref{dipoleexcitation}) yields $N$ vector equations for the $N$ unknown dipole moments ${\bf{p}}^{\left( q \right)}$ (induced by the impinging plane ${\bf{E}}_{0 \bot }$) which, once obtained, allow one to evaluate the reflected and transmitted field, thus completely solving the scattering problem. From a physical point of view Eq.(\ref{dipoleexcitation}) describes the full interaction of the dipole ${\bf{p}}^{\left( q \right)}$ with all the other $N-1$ dipoles through the term containing $G^{\left( {\rm Int} \right)}$ and with the graphene sheet through the terms containing $G_{\rm TE}^{\left( {{\mathop{\rm Re}\nolimits} f} \right)}$ and $G_{\rm TM}^{\left( {{\mathop{\rm Re}\nolimits} f} \right)}$. The direct dipole-dipole interaction is produced by the primary dipole fields and accordingly the tensor $G^{\left( {\rm Int} \right)} \left( {{\bf{r}}_ \bot ^{\left( q \right)}  - {\bf{r}}_ \bot ^{\left( s \right)} } \right)$ only depends on the inter-dipoles distances. On the other hand, the interaction with the graphene sheet is produced by the field emitted by the dipoles lying on the plane $z=-R$ and which is reflected back by graphene to the dipoles, thus accounting for an overall propagation distance equal to $2R$ as evidently shown by the term $e^{ik_{1z} 2R}$ in the tensors $G_{\rm TE}^{\left( {{\mathop{\rm Re}\nolimits} f} \right)}$ and $G_{\rm TM}^{\left( {{\mathop{\rm Re}\nolimits} f} \right)}$. Such term also shows that the dipoles-graphene coupling is dominated by the evanescent waves contribution since the spectral width $\sim 1/(2R)$ is much larger than $k_1$ for the nano-sized spheres with ($R \ll \lambda$) we are considering. As a consequence, graphene plasmon resonances are not ruled out and they effectively play an essential role in the dipole excitation process (see below).

\section{Symmetry analysis}
In order gain physical insight into the dipole excitation process described in the last section, it is worth resorting to the evident symmetry property of the dipoles-graphene system we are considering: it is invariant under a rotation of an angle $2 \pi r/N$ (with $r$ integer) around the $z$- axis. In other words the system has $C_{N}$, the cyclic group of order $N$, as its symmetry group. Consider the N rotation operators
\begin{equation} \label{rotations}
R_r  = \cos \left( {\frac{{2\pi }}{N}r} \right)\left( {{\bf{e}}_x {\bf{e}}_x^T  + {\bf{e}}_y {\bf{e}}_y^T } \right) + \sin \left( {\frac{{2\pi }}{N}r} \right)\left( {{\bf{e}}_y {\bf{e}}_x^T  - {\bf{e}}_x {\bf{e}}_y^T } \right) + {\bf{e}}_z {\bf{e}}_z^T,
\end{equation}
for $r= 0,...,N-1$, which are a representation of the group  $C_N$ on the Euclidean three-dimensional space $\mathcal{E}^3$. From Eqs.(\ref{couplings}) it is straightforward to prove that the three tensors $G$ satisfy the relation
\begin{equation} \label{rotinvariance}
G\left( {R_r {\bf{r}}_ \bot  } \right) = R_r G\left( {{\bf{r}}_ \bot  } \right)R_r^{ - 1}
\end{equation}
which states the rotational invariance of the dipoles interaction.

In order to fully exploit such rotational invariance it is convenient to describe the state of the dipoles ${\bf{p}}^{\left( 0 \right)} , \ldots , {\bf{p}}^{\left( {N - 1} \right)}$ by means of a state vector $\left| {\bf{p}} \right\rangle$ belonging to an abstract $3N$-dimensional Hilbert space $\mathcal{V}$ with the inner product
\begin{equation}
\left\langle {{\bf{a}}}
 \mathrel{\left | {\vphantom {{\bf{a}} {\bf{b}}}}
 \right. \kern-\nulldelimiterspace}
 {{\bf{b}}} \right\rangle  = \sum\limits_{q = 0}^{N - 1} {{\bf{a}}^{\left( q \right)*}  \cdot {\bf{b}}^{\left( q \right)} }.
\end{equation}
The mathematical description is greatly simplified by introducing the $N$ operators ${\hat \Pi ^{\left( q \right)} } : \mathcal{E}^3 \rightarrow \mathcal{V}$, for $q=0,...,N-1$, such that the state $\left| {\bf{a}} \right\rangle  = \hat \Pi ^{\left( r \right)} {\bf{a}}$ corresponds to the dipoles ${\bf{a}}^{\left( q \right)}  = \delta _{q,r} {\bf{a}}$, i.e. they are all vanishing except for the $r$-th which is equal to $\bf a$. The most general state can accordingly be written as
\begin{equation}
\left| {\bf{p}} \right\rangle  = \sum\limits_{q = 0}^{N - 1} {\hat \Pi ^{\left( q \right)} {\bf{p}}^{\left( q \right)} }
\end{equation}
and the operators are easily seen to satisfy the relations
\begin{eqnarray}
 \hat \Pi ^{\left( q \right)\dag } \hat \Pi ^{\left( {q'} \right)}  &=& \delta _{q,q'} I, \nonumber \\
 \sum\limits_{q = 0}^{N - 1} {\hat \Pi ^{\left( q \right)} \hat \Pi ^{\left( q \right)\dag } }  &=& \hat I.
\end{eqnarray}
where $I$ and $\hat I$ are the identities of $\mathcal{E}^3$ and $\mathcal{V}$ respectively. As a consequence of the first of these equations we have ${\bf{p}}^{\left( q \right)}  = \hat \Pi ^{\left( q \right)\dag } \left| {\bf{p}} \right\rangle$.

Using such formalism, the $N$-vector equations of Eq.(\ref{dipoleexcitation}) in $\mathcal{E}^3$ describing the dipoles excitation turn into a single equation in $\mathcal{V}$ which reads
\begin{equation} \label{fundamental}
\left( {\beta  - \hat H} \right)\left| {\bf{p}} \right\rangle  = \left| {{\bf{E}}_0 } \right\rangle
\end{equation}
where
\begin{eqnarray}
 \hat H &=& \sum\limits_{q,s = 0}^{N - 1} {\hat \Pi ^{\left( q \right)} H^{\left( {q,s} \right)} } \hat \Pi ^{\left( s \right)\dag }, \nonumber  \\
 \left| {{\bf{E}}_0 } \right\rangle  &=& \sum\limits_{q = 0}^{N - 1} {\hat \Pi ^{\left( q \right)} } \left[e^{ - ik_1 R} + {e^{ik_1 R} r_{\rm TE} \left( 0 \right)  } \right]\left( {4\pi \varepsilon _0 \varepsilon _1 R^3 } \right){\bf{E}}_{0 \bot }.
\end{eqnarray}
Consider now the $N$ operators $\hat{R}_r$, for $r=0,\ldots,N-1$, acting on $\mathcal{V}$, defined by
\begin{equation}
\hat R_r  = \sum\limits_{q = 0}^{N - 1} {\hat \Pi ^{\left( q \right)} } R_r \hat \Pi ^{\left( {q - r} \right)\dag }
\end{equation}
where the index $(q-r)$ in the right hand side has to be regarded as $(q-r)\: {\rm mod} \: N$, the remainder of $(q-r)/N$. In practice the $q$-th dipole of the state $\hat R_r \left| {\bf{p}} \right\rangle$ is simply equal to the $(q-r)$-th dipole of the state $\left| {\bf{p}} \right\rangle$ rotated by an angle $2 \pi r/N$. It is thus straightforward to prove the relations
\begin{eqnarray}
 \hat R_r^\dag   &=& \hat R_r^{ - 1}, \nonumber  \\
 \hat R_r  &=& \left( {\hat R_1 } \right)^r, \nonumber \\
 \hat R_r \hat R_s  &=& \hat R_{r + s},
\end{eqnarray}
which state that the operators $\hat R_r$ form a unitary representation of $C_{N}$ on $\mathcal{V}$. The characters of such representation are easily evaluated and they are
\begin{equation}
\chi _r  = {\rm Tr} \left( {\hat R_r } \right) = 3N\,\delta _{r,0}
\end{equation}
i.e. they are all vanishing except for the identity operator $\hat R _0$ whose character is $3N$, the dimension of the space $\mathcal{V}$. Now $C_N$ (an abelian group) has exactly $N$ one dimensional irreducible representations: the $r$-th operator of the $n$-th representation ($n= 0,\ldots,N-1$) is simply the multiplication by $e^{-  i\frac{{2\pi }}{N}rn}$ and these are also the characters (the choice of the minus sign corresponds to a specific ordering of the irreducible representations which is here convenient for later developments). Therefore, from standard group theory, the number $m_n$ of times the $n$-th irreducible representation is contained in the representation provided by the operators $\hat{R}_r$ is
\begin{equation}
m_n  = \frac{1}{N}\sum\limits_{r = 0}^{N - 1} {\chi _r } e^{  i\frac{{2\pi }}{N}rn}  = 3.
\end{equation}
Because the representation in unitary, $\mathcal{V}$ is the orthogonal direct sum of $N$ subspaces $\mathcal{V} =  \bigoplus _{n = 0}^{N - 1} \mathcal{V}_n$ each $\mathcal{V}_n$ being spanned by three linearly independent vectors associated to the $n$-th irreducible representation. The vectors $\left| {{\bf{p}}_n } \right\rangle$ belonging to $\mathcal{V}_n$ are eigenvectors of the operators $\hat{R}_r$ with eigenvalue $e^{-  i\frac{{2\pi }}{N}rn}$, i.e.
\begin{equation} \label{irreducible}
\hat R_r \left| {{\bf{p}}_n } \right\rangle  = e^{- i\frac{{2\pi }}{N}rn} \left| {{\bf{p}}_n } \right\rangle,
\end{equation}
which implies
\begin{equation} \label{basis}
{\left| {{\bf{p}}_n } \right\rangle  = \sum\limits_{q = 0}^{N - 1} {e^{i\frac{{2\pi }}{N}qn} \hat \Pi ^{\left( q \right)} R_q {\bf{p}}_n^{\left( 0 \right)} } }
\end{equation}
where ${\bf{p}}_n^{\left( 0 \right)}$ is any three-dimensional vector. As a consequence each vector $\left| {\bf{p}} \right\rangle$ of $\mathcal{V}$ can uniquely be decomposed as
\begin{equation} \label{state}
\left| {\bf{p}} \right\rangle  = \sum\limits_{n = 0}^{N - 1} {\left| {{\bf{p}}_n } \right\rangle }  = \sum\limits_{n = 0}^{N - 1} {\left[ {\sum\limits_{q = 0}^{N - 1} {e^{i\frac{{2\pi }}{N}qn} \hat \Pi ^{\left( q \right)} R_q {\bf{p}}_n^{\left( 0 \right)} } } \right]}.
\end{equation}
Besides the N operators
\begin{equation}
{\hat P_n  = \frac{1}{N}\sum\limits_{q,q' = 0}^{N - 1} {e^{i\frac{{2\pi }}{N}\left( {q - q'} \right)n} \hat \Pi ^{\left( q \right)} R_{q - q'} } \hat \Pi ^{\left( {q'} \right)\dag } },
\end{equation}
for $n=0,...,N-1$, constitute the orthogonal projectors onto the subspaces $\mathcal{V}_n$, i.e.
\begin{eqnarray}
 \hat P_n \hat P_m  &=& \delta _{m,n} \hat P_n, \nonumber  \\
 \sum\limits_{n = 0}^{N - 1} {\hat P_n }  &=& \hat I, \nonumber \\
 \hat P_n \left| {\bf{p}} \right\rangle  &=& \left| {{\bf{p}}_n } \right\rangle.
\end{eqnarray}

Because $C_{N}$ is a symmetry group of the system, Eq.(\ref{rotinvariance}) implies that $\left[ {\hat H,\hat R_r } \right] = 0$ for every $r$ and therefore every subspace $\mathcal{V}_n$ is invariant under $\hat{H}$. Explicitly
\begin{equation}
\hat H\left| {{\bf{p}}_n } \right\rangle  = \sum\limits_{q = 0}^{N - 1} {e^{i\frac{{2\pi }}{N}qn} \hat \Pi ^{\left( q \right)} } R_q \left[ {\sum\limits_{r = 0}^{N - 1} {e^{ - i\frac{{2\pi }}{N}rn} R_{ - r} H^{\left( {r,0} \right)} {\bf{p}}_n^{\left( 0 \right)} } } \right]
\end{equation}
and the independence on $q$ of the three-dimensional vector inside the square brackets proves that ${\hat H\left| {{\bf{p}}_n } \right\rangle } \in \mathcal{V}_n$, or in other words that $\hat P_n \hat H\left| {{\bf{p}}_n } \right\rangle  = \hat H\left| {{\bf{p}}_n } \right\rangle$. As a consequence Eq.(\ref{fundamental}) can be separately solved in each three-dimensional invariant subspace $\mathcal{V}_n$. Therefore, inserting the state of Eq.(\ref{state}) into Eq.(\ref{fundamental}) and projecting the obtained equation onto $\mathcal{V}_n$ we get
\begin{equation}
\left( {\beta  - \hat H} \right)\left| {{\bf{p}}_n } \right\rangle  = \hat P_n \left| {{\bf{E}}_0 } \right\rangle
\end{equation}
that after some algebra yields
\begin{equation} \label{n-rep}
{\left[ {\beta  - \sum\limits_{r = 0}^{N - 1} {e^{ - i\frac{{2\pi }}{N}rn} R_{ - r} H^{\left( {r,0} \right)} } } \right]{\bf{p}}_n^{\left( 0 \right)}  = \left[e^{ - ik_1 R} + {e^{ik_1 R} r_{\rm TE} \left( 0 \right)  } \right]\left( {4\pi \varepsilon _0 \varepsilon _1 R^3 } \right)\left( {\frac{1}{N}\sum\limits_{q = 0}^{N - 1} {e^{ - i\frac{{2\pi }}{N}qn} R_{ - q} } {\bf{E}}_{0 \bot } } \right)}.
\end{equation}
which are N uncoupled vector equations for the N three-dimensional vectors ${{\bf{p}}_n^{\left( 0 \right)} }$. Note that ${{\bf{p}}_n^{\left( 0 \right)} }$ is solely excited by the part of the vector ${\bf{E}}_{0 \bot }$ (the incident plane wave polarization) which `matches' the n-th irreducible representation of $C_N$. Once the vectors ${{\bf{p}}_n^{\left( 0 \right)} }$ are evaluated by inverting the three-dimensional tensors in the left-hand side of Eq.(\ref{n-rep}), the solution of Eqs.(\ref{dipoleexcitation}), from Eq.(\ref{state}), is given by
\begin{equation} \label{dipoles}
{\bf{p}}^{\left( q \right)}  = \hat \Pi ^{\left( q \right)\dag } \left| {\bf{p}} \right\rangle  = \sum\limits_{n = 0}^{N - 1} {e^{i\frac{{2\pi }}{N}qn} R_q {\bf{p}}_n^{\left( 0 \right)} }.
\end{equation}
To summarize, the symmetry analysis has allowed us to decompose the full $3N$-dimensional dipoles excitation process of Eq.(\ref{dipoleexcitation}) into $N$ independent $3$-dimensional problems of Eq.(\ref{n-rep}), each associated to an irreducible representation of $C_N$.

\section{Excitation by circularly polarized light}
The results of the above symmetry analysis show that the incident field selects the specific irreducible representation involved in the dipoles excitation process. The impinging plane wave propagating along the $z-$ axis we are considering has a uniform polarization which is orthogonal to the $z-$axis. In order to investigate the impact of such field distribution on the dipoles excitation, it is convenient to resort to the left and right-hand circular polarization unit vectors
\begin{eqnarray}
 {\bf{ e}}_{\rm L}  &=&  \frac{1}{\sqrt{2}} \left( {\bf e}_x  + i {\bf e}_y \right), \nonumber \\
 {\bf{ e}}_{\rm R}  &=&  \frac{1}{\sqrt{2}} \left( {\bf e}_x  - i {\bf e}_y \right)
\end{eqnarray}
as a basis of the transverse plane, so that
\begin{equation}
{\bf{E}}_{0 \bot }  = E_{0L} {\bf{ e}}_{\rm L}  + E_{0R} {\bf{ e}}_{\rm R}
\end{equation}
and the rotation operators of Eq.(\ref{rotations}) can be conveniently casted as
\begin{equation}
R_r  = e^{ - i\frac{{2\pi }}{N}r} {\bf{ e}}_{\rm L} {\bf{ e}}_{\rm L}^{*T}  + e^{i\frac{{2\pi }}{N}r} {\bf{ e}}_{\rm R} {\bf{ e}}_{\rm R}^{*T}  + {\bf{ e}}_z {\bf{ e}}_z^T.
\end{equation}
Such expressions lead to the relation
\begin{equation}
\frac{1}{N}\sum\limits_{q = 0}^{N - 1} {e^{ - i\frac{{2\pi }}{N}qn} R_{ - q} } {\bf{E}}_{0 \bot }  = \delta _{n,1} E_{\rm 0L} {\bf{ e}}_{\rm L}  + \delta _{n, - 1} E_{\rm 0R} {\bf{ e}}_{\rm R}
\end{equation}
which is an extremely important result for our purposes since, in view of Eq.(\ref{n-rep}), it shows that the impinging plane wave is able to excite only two irreducible representations, namely the $n=1$ and $n=-1$ ones, which are independently excited by the left-hand $E_{\rm 0L}$ and  right-hand $E_{\rm 0R}$ circular components, respectively. It is evident that such two excitation channels are equivalent up to a switch of the left and the right components. Therefore we will hereafter focus on the purely left-polarized impinging plane wave
\begin{equation}
{\bf{E}}_{0 \bot }  = E_0 {\bf{ e}}_{\rm L}.
\end{equation}
for which, Eqs.(\ref{n-rep}) and (\ref{dipoles}) turn into
\begin{eqnarray} \label{EL}
 \left[ {\beta  - \sum\limits_{r = 0}^{N - 1} {e^{ - i\frac{{2\pi }}{N}r} R_{ - r} H^{\left( {r,0} \right)} } } \right]{\bf{p}}^{\left( 0 \right)}  &=& \left[e^{ - ik_1 R} + {e^{ik_1 R} r_{\rm TE} \left( 0 \right)  } \right]\left( {4\pi \varepsilon _0 \varepsilon _1 R^3 } \right)E_0 {\bf{ e}}_{\rm L}, \nonumber  \\
 {\bf{p}}^{\left( q \right)}  &=& e^{i\frac{{2\pi }}{N}q} R_q {\bf{p}}^{\left( 0 \right)},
\end{eqnarray}
where we have dropped the subscript $n$ since the only involved irreducible representation is the $n=1$ one. It is worth noting that, in this situation, the problem of evaluating the dipoles ${\bf{p}}^{\left( 0 \right)},...,{\bf{p}}^{\left( N-1 \right)}$ has been fundamentally reduced to the evaluation of only ${\bf{p}}^{\left( 0 \right)}$, no matter how large is $N$. In fact, the second of Eqs.(\ref{EL}) states that the $q-$th dipole is simply equal to the $0-$th one rotated by angle $2 \pi q /n$ and multiplied by the phase factor  $e^{i\frac{{2\pi }}{N}q}$.

Due to the central role played by the circularly polarized impinging plane wave, it is convenient to also represent the dipoles using the circular basis by setting
\begin{equation}
{\bf{p}}^{\left( q \right)}  = p_{\rm L}^{\left( q \right)} {\bf{ e}}_{\rm L}  + p_{\rm R}^{\left( q \right)} {\bf{ e}}_{\rm R}  + p_z^{\left( q \right)} {\bf{ e}}_z
\end{equation}
so that, after lengthy but straightforward algebra, Eqs.(\ref{EL}) turn into the matrix equations
\begin{eqnarray} \label{p0}
 \left[ {\beta  + M^{\left( {\rm Int} \right)}  + M_{\rm TE}^{\left( \rm Ref \right)} + M_{\rm TM}^{\left( \rm Ref \right)}} \right]\left( {\begin{array}{*{20}c}
   {p_{\rm L}^{\left( 0 \right)} }  \\
   {p_{\rm R}^{\left( 0 \right)} }  \\
   {p_z^{\left( 0 \right)} }  \\
\end{array}} \right) &=& \left[ e^{ - ik_1 R}+ {e^{ik_1 R} r_{\rm TE} \left( 0 \right)  } \right]\left( {4\pi \varepsilon _0 \varepsilon _1 R^3 } \right)E_{0} \left( {\begin{array}{*{20}c}
   1  \\
   0  \\
   0  \\
\end{array}} \right), \nonumber \\
 \left( {\begin{array}{*{20}c}
   {p_{\rm L}^{\left( q \right)} }  \\
   {p_{\rm R}^{\left( q \right)} }  \\
   {p_z^{\left( q \right)} }  \\
\end{array}} \right) &=& \left( {\begin{array}{*{20}c}
   1 & 0 & 0  \\
   0 & {e^{i\frac{{4\pi }}{N}q} } & 0  \\
   0 & 0 & {e^{i\frac{{2\pi }}{N}q} }  \\
\end{array}} \right)\left( {\begin{array}{*{20}c}
   {p_{\rm L}^{\left( 0 \right)} }  \\
   {p_{\rm R}^{\left( 0 \right)} }  \\
   {p_z^{\left( 0 \right)} }  \\
\end{array}} \right),
\end{eqnarray}
where
\begin{eqnarray} \label{M}
 M^{\left( {\rm Int} \right)}  &=& \left( {k_1 R} \right)^3  \sum\limits_{r = 1}^{N - 1} {\left( {\begin{array}{*{20}c}
   { - \left[ {A\left( {k_1 L_r } \right) + B\left( {k_1 L_r } \right)} \right]} & {e^{i\frac{{2\pi }}{N}r} B\left( {k_1 L_r } \right)} & 0  \\
   {e^{i\frac{{2\pi }}{N}r} B\left( {k_1 L_r } \right)} & { - \left[ {A\left( {k_1 L_r } \right) + B\left( {k_1 L_r } \right)} \right]e^{i\frac{{4\pi }}{N}r} } & 0  \\
   0 & 0 & { - A\left( {k_1 L_r } \right)e^{i\frac{{2\pi }}{N}r} }  \\
\end{array}} \right)}, \nonumber  \\
 M_{\rm TE}^{\left( \rm Ref \right)} &=& R^3 \int\limits_0^\infty  {dk_ \bot  } \frac{{e^{i2k_{1z} d} k_ \bot  }}{{2ik_{1z} }}r_{\rm TE} \sum\limits_{r = 0}^{N - 1} {\left( {\begin{array}{*{20}c}
   {k_1^2 J_0 \left( {k_ \bot  L_r } \right)} & { - k_1^2 J_2 \left( {k_ \bot  L_r } \right)e^{i\frac{{2\pi }}{N}r} } & 0  \\
   { - k_1^2 J_2 \left( {k_ \bot  L_r } \right)e^{i\frac{{2\pi }}{N}r} } & {k_1^2 J_0 \left( {k_ \bot  L_r } \right)e^{i\frac{{4\pi }}{N}r} } & 0  \\
   0 & 0 & 0  \\
\end{array}} \right)}, \nonumber  \\
 M_{\rm TM}^{\left( \rm Ref \right)} &=& R^3 \int\limits_0^\infty  {dk_ \bot  } \frac{{e^{i2k_{1z} d} k_ \bot  }}{{2ik_{1z} }}r_{\rm TM} \sum\limits_{r = 0}^{N - 1} {\left( {\begin{array}{*{20}c}
   {k_{1z}^2 J_0 \left( {k_ \bot  L_r } \right)} & {k_{1z}^2 J_2 \left( {k_ \bot  L_r } \right)e^{i\frac{{2\pi }}{N}r} } & { - \sqrt 2 k_ \bot  k_{1z} J_1 \left( {k_ \bot  L_r } \right)e^{i\frac{\pi }{N}r} }  \\
   {k_{1z}^2 J_2 \left( {k_ \bot  L_r } \right)e^{i\frac{{2\pi }}{N}r} } & {k_{1z}^2 J_0 \left( {k_ \bot  L_r } \right)e^{i\frac{{4\pi }}{N}r} } & {\sqrt 2 k_ \bot  k_{1z} J_1 \left( {k_ \bot  L_r } \right)e^{i\frac{{3\pi }}{N}r} }  \\
   { - \sqrt 2 k_ \bot  k_{1z} J_1 \left( {k_ \bot  L_r } \right)e^{i\frac{\pi }{N}r} } & {k_{1z} \sqrt 2 k_ \bot  J_1 \left( {k_ \bot  L_r } \right)e^{i\frac{{3\pi }}{N}r} } & { - 2k_ \bot ^2 J_0 \left( {k_ \bot  L_r } \right)e^{i\frac{{2\pi }}{N}r} }  \\
\end{array}} \right)}, \nonumber  \\
\end{eqnarray}
where $J_n(\xi)$ is the Bessel function of the first kind of order $n$ and
\begin{eqnarray}
 L_r  &=& 2a\sin \left( {\frac{\pi }{N}r} \right), \nonumber \\
 A\left( u \right) &=& \frac{{e^{iu} }}{{u^3 }}\left( { - 1 + iu + u^2 } \right), \nonumber  \\
 B\left( u \right) &=& \frac{{e^{iu} }}{{2u^3 }}\left( {3 - 3iu - u^2 } \right).
\end{eqnarray}
In the derivation of the above expressions, use has been made of the Jacobi-Anger relation
\begin{equation}
e^{i\xi \sin \Phi }  = \sum\limits_{n =  - \infty }^{ + \infty } {J_n \left( \xi  \right)e^{in\Phi } }
\end{equation}
to perform the angular integration in the reciprocal space.

\section{Estimation of the coupling terms}
In order to physically grasp the main features of the dipoles excitation process, it is worth discussing the relative impact of the four contributions to Eq.(\ref{p0}). We here estimate their magnitudes in the situation $R = 30 \:$ nm, $R \ll \lambda$, $a \ll \lambda$ and, for simplicity, in the special case $\varepsilon_1 = \varepsilon_2 = 2.013$. In our analysis we assume that the dielectric permittivity of the sphere is described by the Drude model
\begin{equation}
\varepsilon _{\rm S} \left( \omega  \right) = 1 - \frac{{\omega _{\rm p}^2 }}{{\omega ^2  + i\omega \Gamma }},
\end{equation}
which perfectly applies to transparent conductors with plasma frequency $\omega_{\rm p}$ at mid- and far-infrared angular frequencies.
The frequency $\omega_{\rm S}$ such that
\begin{equation}
\varepsilon _{\rm S} \left( {\omega _{\rm S }} \right) =  - 2\varepsilon _1  + i\varepsilon '_{\rm S}
\end{equation}
labels the single sphere plasmon resonance and we here assume that
\begin{equation}
\varepsilon '_{\rm S}  = {\mathop{\rm Im}\nolimits} \left[ {\varepsilon _{\rm S} \left( {\omega _{\rm S} } \right)} \right] \ll \varepsilon _1.
\end{equation}
In panel $\bf a$ of Fig.2 we plot the real and imaginary parts of $\varepsilon _{\rm S}$ versus the wavelength for a specific $\omega_{\rm p}$ and $\Gamma$ such that $\lambda _{\rm S}  = 2\pi c/\omega _{\rm S} = 3 \: \mu m$ and $\varepsilon '_{\rm p} = 0.1$.

The term $\beta$ is the normalized inverse polarizability of the sphere and it is physically responsible for the excitation of a single sphere. Its normalized absolute value $|\beta|$ is of the order of unity except for wavelengths close to $\lambda _{\rm p}$ (sphere plasmon resonance wavelength) where it achieves a minimum, as clearly shown in panel $\bf b$ of Fig.2 (black line).

The term $M^{\left( {\rm Int} \right)}$ accounts for the direct interaction among the spheres and its leading term can be simply estimated by retaining only the nearest neighbor contributions ($s=1$ and $s=N-1$) so that, for $a \ll \lambda$, the normalized absolute value of its largest matrix element is
\begin{equation}
\left| {M^{\left( {\rm Int} \right)} } \right| \simeq  \left[ {\frac{R}{{2a\sin \left( {\frac{\pi }{N}} \right)}}} \right]^3.
\end{equation}
In other words it scales as the third power of the ratio between the particle radius $R$ and the distance between two adjacent spheres $
{2a\sin \left( {\frac{\pi }{N}} \right)}$ (the edge of the polygon). Note that this contribution is independent on the radiation wavelength $\lambda$ since the condition $a \ll \lambda$ corresponds to a quasi-electrostatic situation where retardation effect are negligible. To maximize this contribution we have considered in our numerical simulation the case where the edge of the polygon is $3$ times larger than the radius $R$ so that we generally assume $\left| {M^{\left( {\rm Int} \right)} } \right| \cong \frac{1}{{27}} = 0.0184$. We have plotted such value in panel $\bf b$ of Fig.2 (red line).

In order to estimate the terms $M_{\rm TE}^{\left( {\rm Re f} \right)}$ and $M_{\rm TM}^{\left( {Ref} \right)}$ it is essential to consider the graphene response and accordingly in panel $\bf c$ and $\bf d$ of Fig.2 we have plotted the real and imaginary parts of normalized graphene surface conductivity $Z_0 \sigma$. Basically at low values of the Fermi energy the graphene response is dominated by absorption since the real part of $Z_0 \sigma$ prevails. For larger values of the Fermi energy, absorption is less important and the imaginary part of $Z_0 \sigma$ increases with the wavelength thus showing a marked metallic behavior.

\begin{figure*} \label{Fig2}
\center
\includegraphics[width=\textwidth]{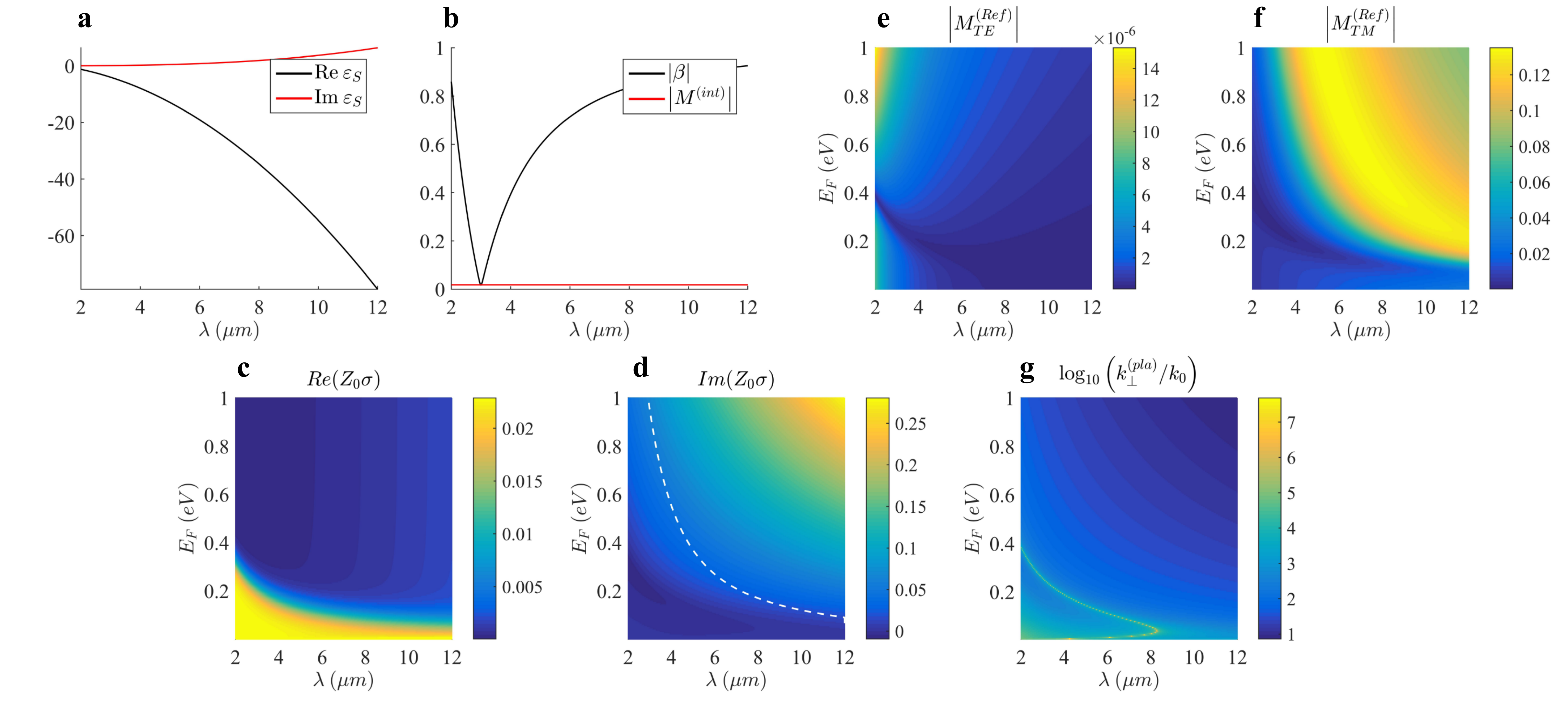}
\caption{{\bf a}. Real and imaginary parts of the spheres permittivity as a function of the wavelength. {\bf b}. Estimations of the single particle and direct dipoles interaction terms. {\bf c-d}. Real and imaginary parts of normalized graphene surface conductivity as functions of the wavelength and the graphene Fermi energy. In panel {\bf d} the curve ${\mathop{\rm Im}\nolimits} \left( {Z_0 \sigma } \right) = \frac{1}{2}\varepsilon _1 k_0 R$ is reported as a white dashed line. The region above and belove such curve correspond to the strong and weak coupling regime, respectively, between the ring and the graphene sheet. {\bf e-f}. Estimations of the TE and TM field coupling terms. {\bf g}. Graphene surface plasmon polariton wavevector $k_\perp^{(\rm pla)}$ normalized with the vacuum wavenumber $k_0$}
\end{figure*}

The term $M_{\rm TE}^{\left( { \rm Ref} \right)}$ is responsible for the coupling of the ring with the graphene sheet produced by the reflected TE field and to get its magnitude we will estimate (for $k_1 R \ll 1$) the integral (see the second of Eqs.(\ref{M}))
\begin{equation}
M_{\rm TE}^{\left( {\rm Ref} \right)}  = R^3 \int\limits_0^\infty  {dk_ \bot  } e^{i2k_{1z} R} \frac{{k_ \bot  k_1^2 }}{{2ik_{1z} }}\left( {\frac{{ - k_0 Z_0 \sigma }}{{2k_{1z}  + k_0 Z_0 \sigma }}} \right)
\end{equation}
which is the leading $s=0$ term of the summation. Splitting the integration into the sum of the homogeneous $k_\bot < k_1$ and evanescent $k_\bot > k_1$ contributions and setting $k_{1z}  = k_1 q$ and $k_{1z}  = ik_1 p$ in the former and the latter, respectively, after some algebra we get
\begin{equation}
M_{\rm TE}^{\left( {\rm Ref} \right)}  = \frac{{i\left( {Z_0 \sigma } \right)\left( {k_1 R} \right)^3 }}{{4 \sqrt{\varepsilon_1} }}\left[ {\int\limits_0^1 {dq} \frac{{e^{i\left( {2k_1 R} \right)q} }}{{q + \left( {\frac{{Z_0 \sigma }}{{2\sqrt {\varepsilon _1 } }}} \right)}} - \int\limits_0^\infty  {dp} \frac{{e^{ - \left( {2k_1 R} \right)p} }}{{p - i\left( {\frac{{Z_0 \sigma }}{{2\sqrt {\varepsilon _1 } }}} \right)}}} \right].
\end{equation}
Since $k_1 R \ll 1$, we can neglect the exponential in the first integral whereas the second one can be expressed in terms of the exponential integral
\begin{equation}
E_1 \left( \zeta  \right) = \int\limits_\zeta ^\infty  {dt} \frac{{e^{ - t} }}{t}, \quad \quad \left(|\arg \zeta| < \pi\right),
\end{equation}
by means of the relation $\int\limits_0^\infty  {dt} \frac{{e^{ - At} }}{{t + B}} = e^{AB} E_1 \left( {AB} \right)$, so that we obtain
\begin{equation}
M_{\rm TE}^{\left( {\rm Ref} \right)}  = \frac{i}{{4 \sqrt{\varepsilon_1} }}\left( {Z_0 \sigma } \right)\left( {k_1 R} \right)^3 \left\{ {\log \left( {1 + \frac{{2\sqrt {\varepsilon _1 } }}{{Z_0 \sigma }}} \right) - \left[ {e^\zeta  E_1 \left( \zeta  \right)} \right]_{\zeta  = \frac{{ - i}}{{\sqrt {\varepsilon _1 } }}\left( {Z_0 \sigma } \right)\left( {k_1 R} \right)} } \right\}
\end{equation}
In the situation we are considering (i.e. $k_1 R \ll 1$ and $|Z_0 \sigma| < 1$ at any wavelength $\lambda$ and any Fermi energy $E_F$), it is evident that the first term is very small. In addition $|\zeta| \ll 1$ also so that, since in this limit $E_1 \left( \zeta  \right) \simeq  - \gamma  - \log \zeta  + \zeta$ (where $\gamma \simeq 0.5772$ is the Euler--Mascheroni constant), the second contribution is very small as well. In panel $\bf e$ of Fig.2 we have plotted $\left| M_{\rm TE}^{\left( {\rm Ref} \right)} \right|$ which results to be much smaller than one at any wavelength $\lambda$ and any Fermi energy $E_F$. This implies that the reflected TE field play a negligible role on the dipoles excitation process and basically this is due to the fact that the TE reflectivity
\begin{equation} \label{rte}
r_{\rm TE} \left( {k_ \bot  } \right) = \frac{{ - Z_0 \sigma }}{{2\sqrt {\varepsilon _1  - \left( {\frac{{k_ \bot  }}{{k_0 }}} \right)^2 }  + Z_0 \sigma }}
\end{equation}
has an absolute value which is generally much smaller than one both for homogeneous ($k_\bot < k_1$) and evanescent ($k_\bot > k_1$) waves. This is also supported by the well known fact that graphene does not effectively admit TE plasmonic resonances.

The term $M_{\rm M}^{\left( {\rm Ref} \right)}$ is responsible for the coupling of the ring with the graphene sheet produced by the reflected TM field and to get its magnitude we will estimate (for $k_1 R \ll 1$) the integral (see the third of Eqs.(\ref{M}))
\begin{equation} \label{MTM}
M_{\rm TM}^{\left( {\rm Ref} \right)}  = R^3 \int\limits_0^\infty  {dk_ \bot  } \frac{{e^{i2k_{1z} R} k_ \bot  k_{1z} }}{{2i}}\left( {\frac{{ - Z_0 \sigma }}{{\frac{{2k_0 \varepsilon _1 }}{{k_{1z} }} + Z_0 \sigma }}} \right)
\end{equation}
which is the leading $s=0$ term of the summation. In analogy with the TE estimation, we again split the integration into the sum of the homogeneous $k_\bot < k_1$ and evanescent $k_\bot > k_1$ contributions and we set $k_{1z}  = k_1 q$ and $k_{1z}  = ik_1 p$ in the former and the latter, respectively, so that we get after some algebra
\begin{equation} \label{MTMetimation}
M_{\rm TM}^{\left( {\rm Ref} \right)}  = \frac{{\left( {k_1 R} \right)^3 }}{2 }\left[ {\frac{{iZ_0 \sigma }}{{2\sqrt {\varepsilon _1 } }}\int\limits_0^1 {dq} \frac{{e^{i\left( {2k_1 R} \right)q} q^3 }}{{1 + \left( {\frac{{Z_0 \sigma }}{{2\sqrt {\varepsilon _1 } }}} \right)q}} - \int\limits_0^\infty  {dp} \frac{{e^{ - \left( {2k_1 R} \right)p} p^3 }}{{p - i\left( {\frac{{2\sqrt {\varepsilon _1 } }}{{Z_0 \sigma }}} \right)}}} \right].
\end{equation}
Since $k_1 R \ll 1$ and $|Z_0 \sigma| < 1$ the integrand of the first integral can be approximated by ${q^3  - \left( {\frac{{Z_0 \sigma }}{{2\sqrt {\varepsilon _1 } }}} \right)q^4 }$ whereas the second integral can again be expressed in term of the exponential integral, thus obtaining
\begin{equation} \label{MTMintegr}
M_{TM}^{\left( {Ref} \right)}  = \frac{{i\left( {Z_0 \sigma } \right)\left( {k_1 R} \right)^3 }}{{8  \sqrt{\varepsilon_1} }}\left( {\frac{1}{2} - \frac{{Z_0 \sigma }}{{5\sqrt {\varepsilon _1 } }}} \right) + \frac{1}{{16 \varepsilon_1}}\left[ {\zeta ^3 e^\zeta  E_1 \left( \zeta  \right) - \zeta ^2  + \zeta  - 2} \right]_{\zeta  = \frac{{4\sqrt {\varepsilon _1 } k_1 R}}{{iZ_0 \sigma }}}.
\end{equation}
The homogeneous waves contribution (first term) is evidently always negligible. On the other hand, the evanescent waves contribution (second term) is generally not negligible since in this case the condition $|\zeta| \ll 1$ is not assured. In panel $\bf f$ of Fig.2 we plot $\left| M_{\rm TM}^{\left( {\rm Ref} \right)} \right|$ from which is evident that the evanescent waves contribution to the TM coupling is relevant only when graphene displays a marked metallic behavior with small absorption. From a physical point of view this behavior is a consequence of the excitation of surface plasmon polaritons (SPPs) on the graphene sheet. Indeed the TM complex reflectivity
\begin{equation}
r_{\rm TM} \left( {k_ \bot  } \right) = \frac{{ - Z_0 \sigma \sqrt {\varepsilon _1  - \left( {\frac{{k_ \bot  }}{{k_0 }}} \right)^2 } }}{{2\varepsilon _1  + Z_0 \sigma \sqrt {\varepsilon _1  - \left( {\frac{{k_ \bot  }}{{k_0 }}} \right)^2 } }}
\end{equation}
is generally not small for evanescent waves and, if ${\rm Im} \left(Z_0 \sigma\right) > 0$, it displays large values around the  SPP transverse wavenumber
\begin{equation}
k_ \bot ^{\left( {\rm pla} \right)}  = k_0 \sqrt {\varepsilon _1  + \left( {\frac{{2\varepsilon _1 }}{{{\mathop{\rm Im}\nolimits} \left( {Z_0 \sigma } \right)}}} \right)^2 }
\end{equation}
where the real part of its denominator vanishes and the plasmonic resonance occurs. In panel $\bf g$ of Fig.2 we have plotted the normalized SPP transverse wavenumber, $k_ \bot ^{\left( {\rm pla} \right)} /k_0$, as function of the both the wavelength and the Fermi energy.
In addition, from the third of Eqs.(\ref{M}) we note that the evanescent spectrum of the TM field is roughly $fe^{ - 2(k_ \bot  R)} (k_ \bot R) ^2$ times the TM reflectivity and the first function has a maximum at $k_{\bot} = 1/R$ and broadening approximately up to $k_{\bot} \simeq 4/R$. Therefore, due to graphene plasmonic resonances, a strong coupling regime occurs if $k_ \bot ^{\left( {\rm pla} \right)}  < 4/R$ which approximately gives
\begin{equation}
{\mathop{\rm Im}\nolimits} \left( {Z_0 \sigma } \right) > \frac{1}{2}\varepsilon _1 k_0 R.
\end{equation}
We have reported in panel $\bf d$ of Fig.2 the curve ${\mathop{\rm Im}\nolimits} \left( {Z_0 \sigma } \right) = \frac{1}{2}\varepsilon _1 k_0 R$ as a dashed white line so that the strong coupling regime occurs on the region lying above this curve. Such region is exactly the one where $\left| M_{\rm TM}^{\left( {\rm Ref} \right)} \right|$ is relevant (see panel $\bf f$ of Fig.2) thus proving the unique role played by graphene SPPs the dipole excitation process.

Roughly, for wavelength $\lambda \lesssim 4 \: \mu$m the SPP wavenumber $k_ \bot ^{\left( {\rm pla} \right)} $ is so large (since ${\rm Im} (Z_0 \sigma)$ is very small) to be effectively out of the TM spectrum and the enhancement of the coupling is not observed. On the other hand, for larger wavelengths, $k_ \bot ^{\left( {\rm pla} \right)} $ decreases (since the imaginary part of $Z_0 \sigma$ increases) and the field is able to trigger plasmonic resonances and to consequently enhance the field coupling. However, this mechanism is effective only for Fermi energies not very small where graphene absorption suppresses the efficiency of the plasmonic resonance, as it is evident from panel $\bf f$ of Fig.2.

\section{Regimes of dipoles excitation}
The estimation of the various terms in the first of Eq.(\ref{p0}) enables to separate the rich phenomenology of the dipoles excitation process into different regimes according to whether the spheres and graphene resonate or not.

{\bf Nonresonant spheres and nonresonant graphene}. Over the region below the dashed white line of panel $\bf d$ of Fig.2 and for wavelengths not close to $\lambda_{\rm S}$, the term $\beta$ is much greater than all the other coupling terms. In fact, in this regime the nonresonant spheres radiate a weak field so that their direct coupling can be neglected and the field reflected by graphene can also be neglected since it is not resonant. Accordingly, after retaining only $\beta$ in the first of Eq.(\ref{p0}) and neglecting the reflectivity $r_{\rm TE} \left( 0 \right)$, we obtain
\begin{eqnarray}
p_{\rm L}^{\left( 0 \right)}  &=& \alpha \left[e^{-i k_1 R} + e^{i k_1 R} r_{\rm TE} (0) \right] E_0, \nonumber  \\
p_{\rm R}^{\left( 0 \right)}  &=& 0, \nonumber \\
p_z^{\left( 0 \right)}  &=& 0,
\end{eqnarray}
evidently stating that the dipoles are not coupled and they respond solely to the imping plane wave.

{\bf Resonant spheres and nonresonant graphene}. In the region below the dashed white line of panel $\bf d$ of Fig.2 and for wavelengths  close to $\lambda_{\rm S}$, the direct coupling term $M^{\left(\rm Int\right)}$ can be comparable with $\beta$ whereas both
$M_{\rm TE}^{\left( \rm Ref \right)}$ and $M_{\rm TM}^{\left( \rm Ref \right)}$ are negligible. Physically, in this regime the resonant spheres radiate a field whose strength is sufficiently large to trigger an efficient direct mutual coupling but the field reflected by nonresonant graphene can be neglected. In order to investigate this regime, we expand $\beta$ up to the first order of $(\omega - \omega_{\rm S})$
\begin{equation} \label{sviluppo}
\beta \left( \omega  \right) \simeq \frac{1}{{\omega_{\rm S} }}\left( {\frac{{2 + 4\varepsilon _1 }}{{3\varepsilon _1 }}} \right)\left( {  \omega _{\rm S}  - \omega   - i\gamma _{\rm S} } \right),
\end{equation}
where $\gamma  = \omega _{\rm S} \frac{{\varepsilon '_{\rm S} }}{{2 + 4\varepsilon _1 }}$ and we resort to the quasi-static approximation $a \ll \lambda$ thus obtaining
\begin{equation}
M^{\left( {\rm int} \right)}  = \frac{1}{2 }\left( {\frac{R}{{2a}}} \right)^3 \sum\limits_{s = 1}^{N - 1} {\left( {\begin{array}{*{20}c}
   { - 1} & {3e^{i\frac{{2\pi }}{N}s} } & 0  \\
   {3e^{i\frac{{2\pi }}{N}s} } & { - e^{i\frac{{4\pi }}{N}s} } & 0  \\
   0 & 0 & {2e^{i\frac{{2\pi }}{N}s} }  \\
\end{array}} \right)\frac{1}{{\sin ^3 \left( {\frac{\pi }{N}s} \right)}}}.
\end{equation}
so that the first of Eq.(\ref{p0}), without $M_{\rm TE}^{\left( \rm Ref \right)}$ and $M_{\rm TM}^{\left( \rm Ref \right)}$, yields
\begin{equation} \label{nographene}
\left( {\begin{array}{*{20}c}
   {\left( {\omega _{\rm S}  + \Delta \omega _{\rm L} } \right) - \omega  - i\gamma } & g & 0  \\
   g & {\left( {\omega _{\rm S}  + \Delta \omega _{\rm R} } \right) - \omega  - i\gamma } & 0  \\
   0 & 0 & {\left( {\omega _{\rm S}  + \Delta \omega _z } \right) - \omega  - i\gamma }  \\
\end{array}} \right)\left( {\begin{array}{*{20}c}
   { p_{\rm L}^{\left( 0 \right)} }  \\
   {p_{\rm R}^{\left( 0 \right)} }  \\
   {p_z^{\left( 0 \right)} }  \\
\end{array}} \right) = \left( {\begin{array}{*{20}c}
   {f_{\rm L} }  \\
   0  \\
   0  \\
\end{array}} \right)
\end{equation}
where
\begin{eqnarray}
 \Delta \omega _{\rm L}  &=& \omega _{\rm S} \left[ { - \frac{3}{2}\left( {\frac{{ \varepsilon_1 }}{{2 + 4\varepsilon _1 }}} \right)\left( {\frac{R}{{2a}}} \right)^3 \sum\limits_{s = 1}^{N - 1} {\frac{1}{{\sin ^3 \left( {\frac{\pi }{N}s} \right)}}} } \right], \nonumber \\
 \Delta \omega _{\rm R}  &=& \omega _{\rm S} \left[ { - \frac{3}{2}\left( {\frac{{\varepsilon_1 }}{{2 + 4\varepsilon _1 }}} \right)\left( {\frac{R}{{2a}}} \right)^3 \sum\limits_{s = 1}^{N - 1} {\frac{{e^{i\frac{{4\pi }}{N}s} }}{{\sin ^3 \left( {\frac{\pi }{N}s} \right)}}} } \right], \nonumber  \\
 \Delta \omega _z  &=& \omega _{\rm S} \left[ {3\left( {\frac{{\varepsilon_1 }}{{2 + 4\varepsilon _1 }}} \right)\left( {\frac{R}{{2a}}} \right)^3 \sum\limits_{s = 1}^{N - 1} {\frac{{e^{i\frac{{2\pi }}{N}s} }}{{\sin ^3 \left( {\frac{\pi }{N}s} \right)}}} } \right], \nonumber  \\
 g &=& \omega _{\rm S} \left[ {\frac{9}{2}\left( {\frac{{\varepsilon_1 }}{{2 + 4\varepsilon _1 }}} \right)\left( {\frac{R}{{2a}}} \right)^3 \sum\limits_{s = 1}^{N - 1} {\frac{{e^{i\frac{{2\pi }}{N}s} }}{{\sin ^3 \left( {\frac{\pi }{N}s} \right)}}} } \right], \nonumber  \\
 f_{\rm L}  &=& \omega _{\rm S} \left[ {3\left( {\frac{{\varepsilon_1 }}{{2 + 4\varepsilon _1 }}} \right) \left[e^{-i k_1 R} + e^{i k_1 R} r_{\rm TE} (0) \right] 4\pi \varepsilon _0 \varepsilon_1 R^3 E_0 } \right].
\end{eqnarray}
Note that $\Delta \omega_{\rm L}$, $\Delta \omega_{\rm R}$, $\Delta \omega_z$ and $g$ are all real quantities. Eq. (\ref{nographene}) states that ${p_{\rm L}^{\left( 0 \right)} }$, ${p_{\rm R}^{\left( 0 \right)} }$ and ${p_z^{\left( 0 \right)} }$ can be viewed as the complex amplitudes of three oscillators, the first of which being driven by the externally applied force $f_{\rm L}$.  Their natural frequencies $\omega_{\rm S} + \Delta \omega_{\rm L}$, $\omega_{\rm S} + \Delta \omega_{\rm R}$ and $\omega_{\rm S} + \Delta \omega_z$ are all slightly shifted from $\omega_{\rm S}$ [since $R <  2a\sin \left( {\frac{\pi }{N}} \right)$] whereas their damping coefficients are all equal to $\gamma$ which is physically provided by spheres absorption. The the first two oscillators are mutually coupled (with coupling coefficient $g$) whereas the third is independent on the others. It is worth noting that equations similar to Eqs.(\ref{nographene}) have already been used in photonics to model Fano resonances and related phenomena \cite{Limon}. From Eq.(\ref{nographene}) we readily obtain
\begin{eqnarray} \label{pnographene}
  p_{\rm L}^{\left( 0 \right)}  &=& \frac{{\left( {\omega _{\rm S}  + \Delta \omega _{\rm R} } \right) - \omega  - i\gamma }}{{\left( {\omega  - \omega _ +   + i\gamma } \right)\left( {\omega  - \omega _ -   + i\gamma } \right)}}f_{\rm L}, \nonumber  \\
  p_{\rm R}^{\left( 0 \right)}  &=&  - \frac{g}{{\left( {\omega  - \omega _ +   + i\gamma } \right)\left( {\omega  - \omega _ -   + i\gamma } \right)}}f_{\rm L}, \nonumber  \\
  p_z^{\left( 0 \right)}  &=& 0,
\end{eqnarray}
where
\begin{eqnarray}
 \omega _ +   &=& \omega _{\rm S}  + \frac{1}{2}\left[ {\left( {\Delta \omega _{\rm L}  + \Delta \omega _{\rm R} } \right) + \sqrt {\left( {\Delta \omega _{\rm L}  - \Delta \omega _{\rm R} } \right)^2  + 4g^2 } } \right], \nonumber \\
 \omega _ -   &=& \omega _{\rm S}  + \frac{1}{2}\left[ {\left( {\Delta \omega _{\rm L}  + \Delta \omega _{\rm R} } \right) - \sqrt {\left( {\Delta \omega _{\rm L}  - \Delta \omega _{\rm R} } \right)^2  + 4g^2 } } \right].
\end{eqnarray}
From Eq.(\ref{pnographene}) we conclude that the dipole longitudinal component $p_z^{\left( 0 \right)}$ is not excited whereas the circular components $p_{\rm L}^{\left( 0 \right)}$ and $p_{\rm R}^{\left( 0 \right)}$ display two resonance peaks at $\omega_+$ and $\omega_-$. Basically these two resonances result from the splitting of the single particle resonance at $\omega_{\rm S}$ produced by the coupling of the two oscillators. In addition, the absolute value of $p_{\rm L}^{\left( 0 \right)}$ attains a minimum at $\omega = \omega_{\rm S} + \Delta \omega_{\rm R}$ (the frequency of the undriven oscillator) which is spectrally located between the two resonance frequencies since $\omega_- < \omega_{\rm S} + \Delta \omega_{\rm R} < \omega_+$. Such phenomenology is a typical example of Fano resonances even though the strict Fano mechanism would require two oscillators with different dampings such that the damping coefficient of the undriven oscillator is much smaller than the other \cite{Limon}.

{\bf Nonresonant spheres and resonant graphene} In the region above the dashed white line of panel $\bf d$ of Fig.2 and for wavelengths  not close to $\lambda_{\rm S}$, the direct coupling term $M^{\left(\rm Int\right)}$ and $M_{\rm TE}^{\left( \rm Ref \right)}$ are negligible whereas $M_{\rm TM}^{\left( \rm Ref \right)}$ is generally comparable with $\beta$. This is a consequence of the fact that the nonresonant spheres radiate a weak field not supporting an efficient direct mutual coupling among them but the TM component of the reflected field, due to the occurrence of graphene plasmonic resonances, is large enough to entail an efficient sphere coupling. To grasp the main features of this regime one can resort to the basic fact that the TM reflectivity $r_{\rm TM}(k_\perp)$ is large only for $k_\perp$ very close to $k_\perp^{(\rm pla)} \gg k_0$ and therefore, in a very crude approximation, one can set $k_{1z}  \cong ik_ \bot$ into the third of Eqs.(\ref{M}) and bring out of the integral the matrix term evaluated at $k_\perp^{(\rm pla)}$, thus obtaining
\begin{equation} \label{MTMapprox}
M_{\rm TM}^{\left( {{\mathop{\rm Re}\nolimits} f} \right)}  \cong \left[ {\frac{{R^3 }}{{2 }}\int\limits_0^\infty  {dk_ \bot  } e^{ - 2k_ \bot  R} k_ \bot ^2 r_{\rm TM} } \right]\sum\limits_{r = 0}^{N - 1} {\left( {\begin{array}{*{20}c}
   {J_0 \left( {k_ \bot ^{\left( {\rm pla} \right)} L_r } \right)} & {J_2 \left( {k_ \bot ^{\left( {\rm pla} \right)} L_r } \right)e^{i\frac{{2\pi }}{N}r} } & {i\sqrt 2 J_1 \left( {k_ \bot ^{\left( {\rm pla} \right)} L_r } \right)e^{i\frac{\pi }{N}r} }  \\
   {J_2 \left( {k_ \bot ^{\left( {\rm pla} \right)} L_r } \right)e^{i\frac{{2\pi }}{N}r} } & {J_0 \left( {k_ \bot ^{\left( {\rm pla} \right)} L_r } \right)e^{i\frac{{4\pi }}{N}r} } & { - i\sqrt 2 J_1 \left( {k_ \bot ^{\left( {\rm pla} \right)} L_r } \right)e^{i\frac{{3\pi }}{N}r} }  \\
   {i\sqrt 2 J_1 \left( {k_ \bot ^{\left( {\rm pla} \right)} L_r } \right)e^{i\frac{\pi }{N}r} } & { - i\sqrt 2 J_1 \left( {k_ \bot ^{\left( {\rm pla} \right)} L_r } \right)e^{i\frac{{3\pi }}{N}r} } & {2J_0 \left( {k_ \bot ^{\left( {\rm pla} \right)} L_r } \right)e^{i\frac{{2\pi }}{N}r} }  \\
\end{array}} \right)}.
\end{equation}
The scalar factor containing the integral is the same we have considered in the above section whereas the matrix is responsible for two main effects. First, the Bessel functions are evaluated at
\begin{equation}
k_ \bot ^{\left( {\rm pla} \right)} L_r  \simeq k_0 a\frac{{4\varepsilon _1 }}{{{\mathop{\rm Im}\nolimits} \left( {Z_0 \sigma } \right)}}\sin \left( {\frac{\pi }{N}r} \right)
\end{equation}
so that, if the spheres are sufficiently packed their plasmon assisted coupling displays oscillations at different Fermi energies or at different wavelengths. This explains why in this regime it is possible to efficiently modulate through electrical gating the dipole excitation, and hence the overall response of the system. Second, all the matrix elements in Eq.(\ref{MTMapprox}) are mutually comparable and multiplied by the same scalar factor, thus implying that the excited dipole will have a generally not vanishing $z$-component. From a physical point of view, such an induced longitudinal dipoles polarization is a consequence of the remarkable component of the electric field perpendicular to the graphene sheet associated to SPPs. This effect will essential for the analysis of the angular momentum carried by the transmitted field (see below). It is worth noting that in the above considered regime where graphene is not resonant, the excited dipoles have a negligible $z$-component so that all the effects we will discuss below based on the longitudinal dipole polarization are entirely due to the occurrence of graphene plasmonic resonances.

{\bf Resonant spheres and resonant graphene} In the region above the dashed white line of panel $\bf d$ of Fig.2 and for wavelengths close to $\lambda_{\rm S}$, all the coupling terms are comparable since the field radiated by the resonant sphere and the field reflected by graphene are both strong. In this regime the oscillating character of the graphene-supported coupling is further enhanced by the direct coupling so that a marked tunability of the system response is achieved through the graphene Fermi energy. Note, incidentally, that the spectral location of the sphere resonance wavelength $\lambda_{\rm S}$ plays here a central role since the graphene optical response dramatically depends on the specific infrared spectral range.

\section{The transmitted field}
The above discussed dipoles excitation process evidently yields a transmitted field (in the half-space $z>0$) whose rich structure is characterized by deep subwavelength features in the near field. These are a consequence of both the spatial location of the dipoles ${\bf{r}}_ \bot ^{\left( s \right)}  = R_s {\bf{r}}_ \bot ^{\left( 0 \right)}$ (from Eq.(\ref{rq})) and of their mutual excitation state ${\bf{p}}^{\left( s \right)}  = e^{i\frac{{2\pi }}{N}s} R_s {\bf{p}}^{\left( 0 \right)}$ (second of Eqs.(\ref{EL})) produced by the left circularly polarized impinging plane wave. Using these relations, the second of Eqs.(\ref{RefTraFields}) supplemented by the spectrum of Eq.(\ref{incidentspectrum}) can be written as
\begin{equation} \label{transmitted}
{\bf{E}}^{\left( \rm T \right)} \left( {\bf{r}} \right) = e^{ik_2 z} t_{\rm TE} \left( 0 \right)E_0 {\bf{ e}}_{\rm L}  + \sum\limits_{s = 0}^{N - 1} {e^{i\frac{{2\pi }}{N}s} } R_s {\bf{E}}^{\left( 0 \right)} \left( {R_s^{ - 1} {\bf{r}}} \right),
\end{equation}
where
\begin{equation} \label{E00}
{{\bf{E}}^{\left( 0 \right)} \left( {\bf{r}} \right) = \int {d^2 {\bf{k}}_ \bot  } e^{i{\bf{k}}_ \bot   \cdot \left( {{\bf{r}}_ \bot   - {\bf{r}}_ \bot ^{\left( 0 \right)} } \right)} e^{ik_{2z} z} \frac{{ie^{ik_{1z} R} }}{{8\pi ^2 \varepsilon _0 \varepsilon _1 }}\left[ {\frac{{k_1^2 }}{{k_{1z} }}t_{\rm TE} \left( {I_ \bot   - \frac{{{\bf{k}}_ \bot  {\bf{k}}_ \bot ^T }}{{k_ \bot ^2 }}} \right) + \frac{{k_ \bot ^2 }}{{k_{1z} }}t_{\rm TM} \left( {\frac{{k_{1z} }}{{k_ \bot ^2 }}{\bf{k}}_ \bot   - \frac{{k_{1z} }}{{k_{2z} }}{\bf{ e}}_z } \right)\left( {\frac{{k_{1z} }}{{k_ \bot ^2 }}{\bf{k}}_ \bot   - {\bf{e}}_z } \right)^T } \right]{\bf{p}}^{\left( 0 \right)} }.
\end{equation}
Note that ${\bf{E}}^{\left( 0 \right)} \left( {\bf{r}} \right)$ is the transmitted field produced by the $0$-th dipole ${\bf{p}}^{\left( 0 \right)}$ located at ${\bf{r}}_ \bot ^{\left( 0 \right)}$, so that Eq.(\ref{transmitted}) clearly shows that the total transmitted field is the superposition of the $N$ dipole fields, weighted by the phase factors ${e^{i\frac{{2\pi }}{N}s} }$, each one amounting to ${\bf{E}}^{\left( 0 \right)} \left( {\bf{r}} \right)$ after a rotation of $2 \pi s/N$ around the $z$-axis. In order to exploit the properties of spatial rotations and to elucidate the field structure, it is convenient to consider the orbital, spin and total angular momentum operators
\begin{eqnarray}
 {\mathbfcal{L}} &=& \frac{1}{i}{\bf{r}} \times \nabla, \nonumber  \\
 {\mathbfcal{S}} &=& \frac{1}{i}{\bf{ e}}_k \left( {\varepsilon _{knm} {\bf{ e}}_n {\bf{ e}}_m^T } \right), \nonumber  \\
 {\mathbfcal{J}} &=& {\mathbfcal{L}} + {\mathbfcal{S}},
\end{eqnarray}
since $\mathbfcal{J} \cdot {\bf n}$ is the infinitesimal generator of rotations around the direction ${\bf n}$. Specifically the operator which rotates a vector field of an angle $2\pi r/N$ around the $z$-axis is
\begin{equation}
\mathcal{R}_r  = e^{ - i\left( {\frac{{2\pi }}{N}r} \right)\mathcal{J}_z },
\end{equation}
where
\begin{equation}
\mathcal{J}_z  = \frac{1}{i}\left( {x\partial _y  - y\partial _x } \right) + \left( {\bf{ e}}_{\rm L} {\bf{ e}}_{\rm L}^{*T}  - {\bf{ e}}_{\rm R} {\bf{ e}}_{\rm R}^{*T} \right),
\end{equation}
so that
\begin{equation}
\mathcal{R}_r {\bf{F}}\left( {\bf{r}} \right) = R_r {\bf{F}}\left( {R_r^{ - 1} {\bf{r}}} \right).
\end{equation}
As a consequence Eq.(\ref{transmitted}) can be casted as
\begin{equation} \label{transmitted2}
{\bf{E}}^{\left( \rm T \right)} \left( {\bf{r}} \right) = e^{ik_2 z} t_{\rm TE} \left( 0 \right)E_0 {\bf{\hat e}}_{\rm L}  + \sum\limits_{s = 0}^{N - 1} {e^{ - i\frac{{2\pi }}{N}s\left( {\mathcal{J}_z  - 1} \right)} } {\bf{E}}^{\left( 0 \right)} \left( {\bf{r}} \right)
\end{equation}
and this expression suggests to switch to the angular momentum representation; due to Fourier representation of Eq.(\ref{E00}), it is convenient to resort to a basis of common eigenvectors of the transverse laplacian and the total angular momentum, i.e.
\begin{eqnarray} \label{eigen}
 \nabla _ \bot ^2 {\bf{U}}_{k_ \bot  ,m}^{\left( q \right)}  &=&  - k_ \bot ^2 {\bf{U}}_{k_ \bot  ,m}^{\left( q \right)} , \nonumber \\
 \mathcal{J}_z {\bf{U}}_{k_ \bot  ,m}^{\left( q \right)}  &=& m{\bf{U}}_{k_ \bot  ,m}^{\left( q \right)} .
\end{eqnarray}
Using polar coordinates ${\bf{r}}_ \bot   = r_ \bot  \left( {\cos \varphi {\bf{e}}_x  + \sin \varphi {\bf{e}}_y } \right)$, it easy to show that these equations are satisfied by the vectors
\begin{eqnarray} \label{UTETM}
 {\bf{U}}_{k_ \bot  ,m}^{\left( {\rm TE} \right)} \left( {{\bf{r}}_ \bot  } \right) &=& \sqrt {\frac{{k_ \bot  }}{{4\pi }}} \left[ {J_{m - 1} \left( {k_ \bot  r_ \bot  } \right)e^{i\left( {m - 1} \right)\varphi } {\bf{ e}}_{\rm L}  + J_{m + 1} \left( {k_ \bot  r_ \bot  } \right)e^{i\left( {m + 1} \right)\varphi } {\bf{ e}}_{\rm R} } \right], \nonumber \\
 {\bf{U}}_{k_ \bot  ,m}^{\left( {\rm TM} \right)} \left( {{\bf{r}}_ \bot  } \right) &=& \sqrt {\frac{{k_ \bot  }}{{4\pi }}} \left[ {J_{m - 1} \left( {k_ \bot  r_ \bot  } \right)e^{i\left( {m - 1} \right)\varphi } {\bf{ e}}_{\rm L}  - J_{m + 1} \left( {k_ \bot  r_ \bot  } \right)e^{i\left( {m + 1} \right)\varphi } {\bf{ e}}_{\rm R}  - i\sqrt 2 \frac{{k_ \bot  }}{{k_{2z} }}J_m \left( {k_ \bot  r_ \bot  } \right)e^{im\varphi } {\bf{ e}}_z } \right],
\end{eqnarray}
whose TE and TM character is a consequence of the relations
\begin{eqnarray}
 \nabla _ \bot  \nabla _ \bot   \cdot \left[ {{\bf{U}}_{k_ \bot  ,m}^{\left( {\rm TE} \right)} } \right]_ \bot   &=& 0, \nonumber \\
 \nabla _ \bot  \nabla _ \bot   \cdot \left[ {{\bf{U}}_{k_ \bot  ,m}^{\left( {\rm TM} \right)} } \right]_ \bot   &=&  - k_ \bot ^2 \left[ {{\bf{U}}_{k_ \bot  ,m}^{\left( {\rm TM} \right)} } \right]_ \bot.
\end{eqnarray}
It is also easy to show that such eigenvectors satisfy the orthogonality relation
\begin{equation} \label{orthog}
\int {d^2 {\bf{r}}_ \bot  } \left[ {{\bf{U}}_{k_ \bot  ,m}^{\left( q \right)} \left( {{\bf{r}}_ \bot  } \right)} \right]^{*T} \left[ {{\bf{U}}_{k'_ \bot  ,m'}^{\left( {q'} \right)} \left( {{\bf{r}}_ \bot  } \right)} \right] = \left[ {1 + \frac{{k_ \bot ^2 }}{{\left| {k_{2z} } \right|^2 }}\delta _{q,{\rm TM}} } \right]\delta \left( {k_ \bot   - k'_ \bot  } \right)\delta _{m,m'} \delta _{qq'}.
\end{equation}
After a somehow lengthy but straightforward calculation, it is possible to prove from Eq.(\ref{E00}) that the transmitted field of the $0$-th dipole has the angular momentum representation
\begin{equation} \label{E000}
{\bf{E}}^{\left( 0 \right)} \left( {\bf{r}} \right) = \frac{1}{N}\sum\limits_{m =  - \infty }^{ + \infty } {\int\limits_0^\infty  {dk_ \bot  } e^{ik_{2z} z} \left[ {b_{k_ \bot  ,m}^{\left( {\rm TE} \right)} {\bf{U}}_{k_ \bot  ,m}^{\left( {\rm TE} \right)} \left( {{\bf{r}}_ \bot  } \right) + b_{k_ \bot  ,m}^{\left( {\rm TM} \right)} {\bf{U}}_{k_ \bot  ,m}^{\left( {\rm TM} \right)} \left( {{\bf{r}}_ \bot  } \right)} \right]}
\end{equation}
where
\begin{eqnarray} \label{bbb}
 b_{k_ \bot  ,m}^{\left( {\rm TE} \right)}  &=& \frac{{iN}}{{4\sqrt \pi  \varepsilon _0 \varepsilon _1 }}e^{ik_{1z} R} \sqrt {k_ \bot  } \frac{{k_1^2  }}{{k_{1z} }}  t_{\rm TE} \left[ {J_{m - 1} \left( {k_ \bot  a} \right)p_{\rm L}^{\left( 0 \right)}  + J_{m + 1} \left( {k_ \bot  a} \right)p_{\rm R}^{\left( 0 \right)} } \right], \nonumber \\
 b_{k_ \bot  ,m}^{\left( {\rm TM} \right)}  &=& \frac{{iN}}{{4\sqrt \pi  \varepsilon _0 \varepsilon _1 }}e^{ik_{1z} R}  \sqrt {k_ \bot  } k_{1z} t_{\rm TM} \left[ {J_{m - 1} \left( {k_ \bot  a} \right)p_{\rm L}^{\left( 0 \right)}  - J_{m + 1} \left( {k_ \bot  a} \right)p_{\rm R}^{\left( 0 \right)}  + i\sqrt 2 \frac{{k_ \bot  }}{{k_{1z} }}J_m \left( {k_ \bot  a} \right)p_z^{\left( 0 \right)} } \right].
\end{eqnarray}
Note that ${\bf{E}}^{\left( 0 \right)} \left( {\bf{r}} \right) $ turns out to be the superposition of all the possible angular momentum eigenstates and this is due to the fact that the orbital angular momentum $\mathcal{L}_z$ is defined with respect the polygon axis (the $z$-axis) whereas the $0$-th dipole is off-axis located at ${\bf{r}}_ \bot ^{\left( 0 \right)} = a {\bf e}_x$. Accordingly, in the limit case of $a=0$, only the eigenstates with $m=1,0,-1$ are involved and they are excited by $p_{\rm L}^{\left( 0 \right)}$, $p_{\rm R}^{\left( 0 \right)}$ and $p_z^{\left( 0 \right)}$, respectively. Inserting Eq.(\ref{E000}) into Eq.(\ref{transmitted2}), after using the second of Eqs.(\ref{eigen}) and the mathematical identity
\begin{equation}
\frac{1}{N}\sum\limits_{s = 0}^{N - 1} {e^{ - i\frac{{2\pi }}{N}qs} }  = \sum\limits_{n =  - \infty }^{ + \infty } {\delta _{q,nN} },
\end{equation}
we get the angular momentum representation of the overall transmitted field
\begin{equation} \label{OAMrep}
{\bf{E}}^{\left( \rm T \right)} \left( {\bf{r}} \right) = e^{ik_2 z} t_{\rm TE} \left( 0 \right)E_0 {\bf{\hat e}}_{\rm L}  + \sum\limits_{n =  - \infty }^{ + \infty } {\int\limits_0^\infty  {dk_ \bot  } e^{ik_{2z} z} \left[ {b_{k_ \bot  ,nN + 1}^{\left( {\rm TE} \right)} {\bf{U}}_{k_ \bot  ,nN + 1}^{\left( {\rm TE} \right)} \left( {{\bf{r}}_ \bot  } \right) + b_{k_ \bot  ,nN + 1}^{\left( {\rm TM} \right)} {\bf{U}}_{k_ \bot  ,nN + 1}^{\left( {\rm TM} \right)} \left( {{\bf{r}}_ \bot  } \right)} \right]},
\end{equation}
revealing that only the angular momentum eigenstates with $m = nM +1$ are involved. From a physical point of view, all the other eigenstates disappear by destructive interference in the superposition of the $N$ dipole fields of Eq.(\ref{transmitted}). On the other hand, the phase factors ${e^{i\frac{{2\pi }}{N}s} }$ in this superposition, originally due to the dipoles excitation by the impinging plane wave whose angular momentum is equal to 1 (left hand polarization), is accordingly responsible for the unit shift of the angular momentum eigenvalues $m = nM +1$. From a geometrical point of view this is related that the fact that
\begin{equation}
 \mathcal{R}_r {\bf{E}}^{\left( \rm T \right)} \left( {\bf{r}} \right) = e^{ - i {\frac{{2\pi }}{N}r} } {\bf{E}}^{\left( \rm T \right)} \left( {\bf{r}} \right)
\end{equation}
which clearly shows that the transmitted electric field belongs to the same irreducible representation of $C_N$ (in the Hilbert space of vector fields) of the dipoles state (see Eq.(\ref{irreducible}) with $n=1$).

\section{Angular momentum steering}
The angular momentum representation of Eq.(\ref{OAMrep}) elucidates the structure of the transmitted field and, interestingly, the expansion coefficients of Eqs.(\ref{bbb}) unexpectedly display a particularly simple structure since they are linear functionals of the $0$-th dipole ${\bf p}^{(0)}$. This evidently implies that the distribution of the angular momentum eigenvalues of the transmitted field depends on the dipoles polarization state or, in other words, that the angular momentum carried by the transmitted field can effectively be tuned through the wavelength and the graphene Fermi energy. Since the impinging plane wave has angular momentum equal to one (left circular polarization) we conclude that the system effectively shows an angular momentum steering functionality. In order to discuss this important point, the evaluation of the angular momentum carried by the transmitted field is in order.

It is well kwnon that the energy and angular momentum densities of a monochromatic field in a transparent and nondispersive medium of dielectric permittivity $\epsilon_2$ are \cite{Blio1}
\begin{eqnarray} \label{UJ}
 U &=& \frac{{\varepsilon _2 }}{4}{\mathop{\rm Re}\nolimits} \left( {\varepsilon _0 {\bf{E}}^{*T} {\bf{E}} + \frac{{\mu _0 }}{{\varepsilon _2 }}{\bf{H}}^{*T} {\bf{H}}} \right), \nonumber \\
 {\bf{J}} &=& \frac{1}{{4\omega }}{\mathop{\rm Re}\nolimits} \left[ {\varepsilon _0 {\bf{E}}^{*T} \left( \mathbfcal{J} \right){\bf{E}} + \frac{{\mu _0 }}{{\varepsilon _2 }}{\bf{H}}^{*T} \left( \mathbfcal{J} \right){\bf{H}}} \right],
\end{eqnarray}
where we have used the notation ${\bf{F}}^{*T} \left( \mathbfcal{O} \right){\bf{F}} = \left( F_i^*  \mathcal{O}_j F_i \right) {\bf{e}}_j$. The above expression for the angular momentum density is strictly equivalent to the standard one ${\bf{J}} = {\bf{r}} \times {\bf{P}}$, where ${\bf{P}}$ is the linear momentum density, and we have chosen it since it is particularly suitable for our analysis. As a matter of fact, such angular momentum density can be derived starting from the Belinfante's decomposition of the linear momentum density \cite{Berry}
\begin{equation} \label{belinfante}
{\bf{P}} = \frac{1}{{2c^2 }}{\mathop{\rm Re}\nolimits} \left( {{\bf{E}} \times {\bf{H}}^* } \right) = {\bf{P}}^{\rm o}  + {\bf{P}}^{\rm s}
\end{equation}
where
\begin{eqnarray}
 {\bf{P}}^{\rm o}  &=& \frac{1}{{4\omega }}{\mathop{\rm Re}\nolimits} \left[ {\varepsilon _0 {\bf{E}}^{*T} \left( {\frac{1}{i}\nabla } \right){\bf{E}} + \frac{{\mu _0 }}{{\varepsilon _2 }}{\bf{H}}^{*T} \left( {\frac{1}{i}\nabla } \right){\bf{H}}} \right], \nonumber \\
{\bf{P}}^{\rm s}  &=& \frac{1}{2}\nabla\times \bf{S}, \nonumber \\
\bf{S} &=&  {\frac{1}{{4\omega }}{\mathop{\rm Re}\nolimits} \left[ {\varepsilon _0 {\bf{E}}^{*T} \left( \mathbfcal{S}  \right){\bf{E}} + \frac{{\mu _0 }}{{\varepsilon _2 }}{\bf{H}}^{*T} \left( \mathbfcal{S} \right){\bf{H}}} \right]}
\end{eqnarray}
are the orbital and spin linear momentum densities, respectively. The orbital contribution ${\bf{P}}^{\rm o}$ is equal to the canonical momentum density and it is proportional to the local phase gradient whereas the spin contribution ${\bf{P}}^{\rm s}$ is the curl of the spin density which has been shown to provide a transverse spin contribution in the presence of evanescent waves \cite{Blio2}.

To investigate the angular momentum carried by the transmitted field, we consider the quantity
\begin{equation} \label{avJz}
\left\langle {J_z } \right\rangle \left( z \right) =  \omega \varepsilon _2 \frac{\displaystyle {\int {d^2 } {\bf{r}}_ \bot  J_z \left( {\bf{r}} \right)}}{\displaystyle {\int {d^2 {\bf{r}}_ \bot  } U\left( {\bf{r}} \right)}} = \frac{\displaystyle {\int {d^2 } {\bf{r}}_ \bot  {\mathop{\rm Re}\nolimits} \left( {\varepsilon _0 {\bf{E}}^{*T} \mathcal{J}_z {\bf{E}} + \frac{{\mu _0 }}{{\varepsilon _2 }} {{\bf{H}}^{*T} } \mathcal{J}_z {\bf{H}}} \right)}}{\displaystyle {\int {d^2 {\bf{r}}_ \bot  } {\mathop{\rm Re}\nolimits} \left( {\varepsilon _0 {\bf{E}}^{*T} {\bf{E}} + \frac{{\mu _0 }}{{\varepsilon _2 }}{\bf{H}}^{*T} {\bf{H}}} \right)}}
\end{equation}
which provides a dimensionless average value of the longitudinal angular momentum density $J_z$ on the plane $z$. Since the plane wave appearing in the Eq.(\ref{OAMrep}) has infinite energy and it does not accounts for the angular momentum exchange upon scattering from the spheres, we will evaluate the average of $J_z$ on the field
\begin{equation} \label{EEE}
{\bf{E}}\left( {\bf{r}} \right) = \sum\limits_{n =  - \infty }^{ + \infty } {\int\limits_0^\infty  {dk_ \bot  } e^{ik_{2z} z} \left[ {b_{k_ \bot  ,nN + 1}^{\left( {\rm TE} \right)} {\bf{U}}_{k_ \bot  ,nN + 1}^{\left( {\rm TE} \right)} \left( {{\bf{r}}_ \bot  } \right) + b_{k_ \bot  ,nN + 1}^{\left( {\rm TM} \right)} {\bf{U}}_{k_ \bot  ,nN + 1}^{\left( {\rm TM} \right)} \left( {{\bf{r}}_ \bot  } \right)} \right]}
\end{equation}
which is properly the field scattered by the dipoles. The magnetic field associated with this field can easily be evaluated from the Maxwell equation $\nabla  \times {\bf{E}} = i\omega \mu _0 {\bf{H}}$ and it is given by
\begin{equation} \label{HHH}
{\bf{H}}\left( {\bf{r}} \right) = \frac{1}{i}\sqrt {\frac{{\varepsilon _0 \varepsilon _2 }}{{\mu _0 }}} \sum\limits_{n =  - \infty }^{ + \infty } {\int\limits_0^\infty  {dk_ \bot  } e^{ik_{2z} z} \left\{ {\left[ {\frac{{k_2 }}{{k_{2z} }}b_{k_ \bot  ,nN + 1}^{\left( {\rm TM} \right)} } \right]{\bf{U}}_{k_ \bot  ,nN + 1}^{\left( {\rm TE} \right)} \left( {{\bf{r}}_ \bot  } \right) + \left[ {\frac{{k_{2z} }}{{k_2 }}b_{k_ \bot  ,nN + 1}^{\left( {\rm TE} \right)} } \right]{\bf{U}}_{k_ \bot  ,nN + 1}^{\left( {\rm TM} \right)} \left( {{\bf{r}}_ \bot  } \right)} \right\}}
\end{equation}
where we have used the relation (also easy to prove)
\begin{eqnarray}
 \nabla  \times \left[ {e^{ik_{2z} z} {\bf{U}}_{k_ \bot  ,m}^{\left( {\rm TE} \right)} } \right] &=& k_{2z} e^{ik_{2z} z} {\bf{U}}_{k_ \bot  ,m}^{\left( {\rm TM} \right)}, \nonumber  \\
 \nabla  \times \left[ {e^{ik_{2z} z} {\bf{U}}_{k_ \bot  ,m}^{\left( {\rm TM} \right)} } \right] &=& \frac{{k_2^2 }}{{k_{2z} }}e^{ik_{2z} z} {\bf{U}}_{k_ \bot  ,m}^{\left( {\rm TE} \right)}.
\end{eqnarray}
Inserting Eqs.(\ref{EEE}) and (\ref{HHH}) into Eq.(\ref{avJz}), after exploiting the orthogonality of the angular momentum eigenstates (see Eq.(\ref{orthog})), we get
\begin{equation} \label{avJz1}
\left\langle {J_z } \right\rangle  = \frac{1}{{W}}\sum\limits_{n =  - \infty }^{ + \infty } {\left( {nN + 1} \right)w_n }
\end{equation}
where
\begin{eqnarray} \label{w}
w_n(z)  &=& \frac{{\varepsilon _0 \varepsilon _2 }}{4}\int\limits_0^\infty  {dk_ \bot  } e^{ - 2{\mathop{\rm Im}\nolimits} \left( {k_{2z} } \right)z} \left( {\left| {k_{2z} } \right|^2  + k_2^2  + k_ \bot ^2 } \right)\left[ {\frac{1}{{k_2^2 }}\left| {b_{k_ \bot  ,nN + 1}^{\left( {\rm TE} \right)} } \right|^2  + \frac{1}{{\left| {k_{2z} } \right|^2 }}\left| {b_{k_ \bot  ,nN + 1}^{\left( {\rm TM} \right)} } \right|^2 } \right], \nonumber \\
W\left( z \right) &=& \sum\limits_{n =  - \infty }^{ + \infty } {w_n \left( z \right)},
\end{eqnarray}
effectively stating that $\left\langle {J_z } \right\rangle$ is the weighted mean of the angular momentum eigenvalues $m = nN+1$ with positive weights $w_n$.

Note that the average of $J_z$ of Eq.(\ref{avJz1}) can be written as
\begin{equation} \label{avJz2}
\left\langle {J_z } \right\rangle  = 1 + \frac{N}{W}\sum\limits_{n =  - \infty }^{ + \infty } {n w_n^{\left(\rm  A \right)} },
\end{equation}
where $w_n^{\left( \rm A \right)}  = \frac{1}{2}\left( {w_n  - w_{ - n} } \right)$ is the antisymmetric part of the sequence $w_n$, this implying that the transmitted angular momentum is equal to one if $w_n$ is even. Now from Eqs.(\ref{bbb}) we get
\begin{eqnarray} \label{bbb1}
 b_{k_ \bot  ,nN + 1}^{\left( {\rm TE} \right)}  &=& \frac{{iN}}{{4\sqrt \pi  \varepsilon _0 \varepsilon _1 }}e^{ik_{1z} R} \frac{{k_1^2 \sqrt {k_ \bot  } }}{{k_{1z} }}t_{\rm TE} \left[ {J_{nN} \left( {k_ \bot  a} \right)p_{\rm L}^{\left( 0 \right)}  + J_{nN + 2} \left( {k_ \bot  a} \right)p_{\rm R}^{\left( 0 \right)} } \right], \nonumber \\
 b_{k_ \bot  ,nN + 1}^{\left( {\rm TM} \right)}  &=& \frac{{iN}}{{4\sqrt \pi  \varepsilon _0 \varepsilon _1 }}e^{ik_{1z} R} k_{1z} \sqrt {k_ \bot  } t_{\rm TM} \left[ {J_{nN} \left( {k_ \bot  a} \right)p_{\rm L}^{\left( 0 \right)}  - J_{nN + 2} \left( {k_ \bot  a} \right)p_{\rm R}^{\left( 0 \right)}  + i\sqrt 2 \frac{{k_ \bot  }}{{k_{1z} }}J_{nN + 1} \left( {k_ \bot  a} \right)p_z^{\left( 0 \right)} } \right],
\end{eqnarray}
from which we deduce that $w_n$ is even only if $p_{\rm R}^{\left( 0 \right)} = p_z^{\left( 0 \right)} =0$ (since $J_{-m} (\xi)= (-1)^m J_{m} (\xi)$). In other words, from a physical point of view the scattered angular momentum is equal to one only if the dipoles are left circularly polarized, as expected. This happens only in the regime where neither the spheres nor graphene are resonant (see Section V) so that we expect a marked exchange of angular momentum in the resonant situations, thus proving the above observation that the angular momentum carried by the scattered field can be manipulated by the dipole excitation process.

\section{Excitation of modes with definite rotation direction}
The above discussion concerning the angular momentum steering have the remarkable consequence that scattered fields with positive or negative angular momentum corresponds to the excitation of graphene surface modes with definite counterclockwise or clockwise rotation direction. In order to get more physical insight into such novel phenomenon we now investigate the properties of the antisymmetric part of the sequence $w_n$.

The first key observation is that, in the regime of resonant graphene, one can neglect the contribution of the TE modes in the expression for $w_n$ in the first of Eqs.(\ref{w}) since graphene resonance is a purely TM phenomenon. In addition, since the largest contribution to the integral comes from the transverse wave vectors $k_\perp$ close to $k_\perp^{(\rm pla)} \gg k_0$, it is possible to set $k_{1z}  \simeq k_{2z}\simeq ik_\bot$ an to restrict the integral to the evanescent part thus getting
\begin{equation} \label{wwnn}
w_n \left( 0 \right) \cong \frac{{N^2 \varepsilon _2 }}{{32\pi \varepsilon _0 \varepsilon _1^2 }}\int\limits_{k_2 }^\infty  {dk_ \bot  } e^{ - 2k_ \bot  R} k_ \bot ^3 \left| {t_{\rm TM} } \right|^2 C_n \left( {k_ \bot  a} \right)
\end{equation}
where
\begin{equation}
C_n \left( \xi  \right) = \left| {J_{nN} \left( \xi  \right)p_{\rm L}^{\left( 0 \right)}  - J_{nN + 2} \left( \xi  \right)p_{\rm R}^{\left( 0 \right)}  + \sqrt 2 J_{nN + 1} \left( \xi  \right)p_z^{\left( 0 \right)} } \right|^2.
\end{equation}
In order to extract the antisymmetric part of $C_n$ it is convenient to transform the Bessel functions in such a way that all their indices are equal. With the help of the well known relations
\begin{eqnarray}
 J_{m - 1} \left( \xi  \right) + J_{m + 1} \left( \xi  \right) &=& \frac{{2m}}{\xi }J_m \left( \xi  \right), \nonumber \\
 J_{m - 1} \left( \xi  \right) - J_{m + 1} \left( \xi  \right) &=& 2J'_m \left( \xi  \right), \nonumber \\
 J_{m + 1} \left( \xi  \right) &=& \frac{m}{\xi }J_m \left( \xi  \right) - J'_m \left( \xi  \right),
\end{eqnarray}
after some algebra we get
\begin{eqnarray}
 C_n^{\left( \rm A \right)}  &=& \frac{1}{2}\left( {C_n  - C_{ - n} } \right) = \nonumber  \\
 & = & - 4nN{\mathop{\rm Re}\nolimits} \left\{ {\left[ {J''_{nN} \left( \xi  \right)p_x^{\left( 0 \right)*}  + J'_{nN} \left( \xi  \right)\left( {\frac{1}{\xi }ip_y^{\left( 0 \right)*}  + p_z^{\left( 0 \right)*} } \right)} \right]\left[ {\left( {\frac{{J'_{nN} \left( \xi  \right)}}{\xi } - \frac{{J_{nN} \left( \xi  \right)}}{{\xi ^2 }}} \right)\left( {p_x^{\left( 0 \right)}  + ip_y^{\left( 0 \right)} } \right) + \frac{{J_{nN} \left( \xi  \right)}}{\xi }p_z^{\left( 0 \right)} } \right]} \right\} + \nonumber  \\
 & +& 4n^3 N^3 \left\{ {\left( { - \frac{{J_{nN}^2 \left( \xi  \right)}}{{\xi ^4 }} + \frac{{J_{nN} \left( \xi  \right)J'_{nN} \left( \xi  \right)}}{{\xi ^3 }}} \right){\mathop{\rm Re}\nolimits} \left[ {i\left( {p_x^{\left( 0 \right)}  + ip_y^{\left( 0 \right)} } \right)p_y^{\left( 0 \right)*} } \right] + \frac{{J_{nN}^2 \left( \xi  \right)}}{{\xi ^3 }}{\mathop{\rm Re}\nolimits} \left( {ip_z^{\left( 0 \right)} p_y^{\left( 0 \right)*} } \right)} \right\}
\end{eqnarray}
where the cartesian components $p_x^{\left( 0 \right)}  = \left( {p_{\rm L}^{\left( 0 \right)}  + p_{\rm R}^{\left( 0 \right)} } \right)/\sqrt{2}$ and $p_y^{\left( 0 \right)}  = i\left( {p_{\rm L}^{\left( 0 \right)}  - p_{\rm R}^{\left( 0 \right)} } \right)/\sqrt{2}$ of the $0$-th dipole have been introduced. Note that the terms inside the curly brackets are even under the inversion $n \rightarrow -n$. The first term proportional to $nN$ can be generally neglected with respect to the second which is proportional to $n^3 N^3$ so that, extracting the antisymmetric part of  Eq.(\ref{wwnn}) and performing the change of variables $\xi  = k_ \bot  a$, we get
\begin{eqnarray}
 w_n^{\left( \rm A \right)} \left( 0 \right) &\cong& n^3 \left[ {\frac{{N^5 \varepsilon _2 }}{{8\pi \varepsilon _0 a^4 \varepsilon _1^2 }}\int\limits_{k_2 a}^\infty  {d\xi } e^{ - \left( {\frac{{2R}}{a}} \right)\xi } \left| {t_{\rm TM} } \right|^2 \left( { - \frac{1}{\xi }J_{nN}^2 \left( \xi  \right) + J_{nN} \left( \xi  \right)J'_{nN} \left( \xi  \right)} \right)} \right]{\mathop{\rm Re}\nolimits} \left[ {i\left( {p_x^{\left( 0 \right)}  + ip_y^{\left( 0 \right)} } \right)p_y^{\left( 0 \right)*} } \right] + \nonumber  \\
 &&+ n^3 \left[ {\frac{{N^5 \varepsilon _2 }}{{8\pi \varepsilon _0 a^4 \varepsilon _1^2 }}\int\limits_{k_2 a}^\infty  {d\xi } e^{ - \left( {\frac{{2R}}{a}} \right)\xi } \left| {t_{\rm TM} } \right|^2 J_{nN}^2 \left( \xi  \right)} \right]{\mathop{\rm Re}\nolimits} \left( {ip_z^{\left( 0 \right)} p_y^{\left( 0 \right)*} } \right).
\end{eqnarray}
Now the plasmonic resonance occurs for $\xi$ of the order of $10$ so that the integral containing $J_{nN}^2 \left( \xi  \right)/\xi$ can be discarded. In addition the integral containing ${J_{nN} \left( \xi  \right)J'_{nN} \left( \xi  \right)}$ can also be neglected due to the positive and negative oscillations of this function. Inserting the residual second term into Eq.(\ref{avJz2}) we get
\begin{equation} \label{avJz4}
\left\langle {J_z } \right\rangle  = 1 + \gamma {\mathop{\rm Re}\nolimits} \left( {ip_z^{\left( 0 \right)} p_y^{\left( 0 \right)*} } \right)
\end{equation}
where
\begin{equation}
\gamma  = \frac{1}{W}\sum\limits_{n =  - \infty }^{ + \infty } {n^4 } \left[ {\frac{{N^6 \varepsilon _2 }}{{8\pi \varepsilon _0 a^4 \varepsilon _1^2 }}\int\limits_{k_2 a}^\infty  {d\xi } e^{ - \left( {\frac{{2R}}{a}} \right)\xi } \left| {t_{\rm TM} } \right|^2 J_{nN}^2 \left( \xi  \right)} \right]
\end{equation}
is a real and positive quantity not depending on the dipole ${\bf p}^{(0)}$. Equation (\ref{avJz4}) is one of the main result of our investigation and it basically reveals that the scattered angular momentum is fundamentally driven by the quantity ${\mathop{\rm Re}\nolimits} \left( {ip_z^{\left( 0 \right)} p_y^{\left( 0 \right)*} } \right)$ associated to the dipole state and whose sign dictates  whether or not the scattered angular momentum exceeds the angular momentum of the impinging plane wave.

\begin{figure*} \label{Fig3}
\center
\includegraphics[width=0.9\textwidth]{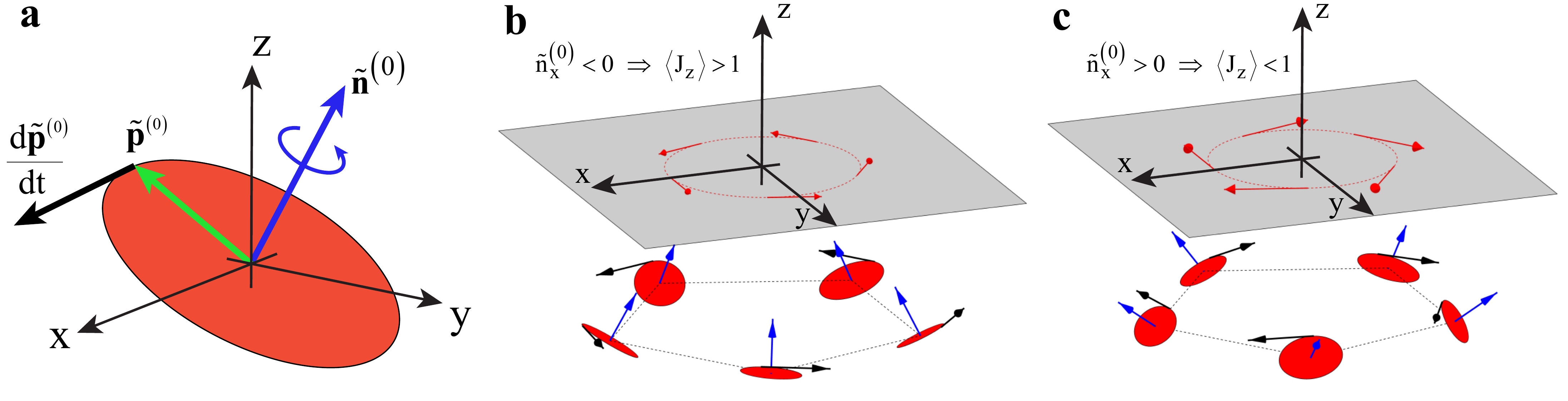}
\caption{\textbf{a} Geometry of the polarization ellipse and of the angular vector ${\bf{\tilde n}}^{\left( 0 \right)}$. (\textbf{b}, \textbf{c}) Mechanism of excitation of surface modes with definite rotation direction.}
\end{figure*}

Equation (\ref{avJz4}) admits a particularly clear physical interpretation related to the meaning of the quantity ${\mathop{\rm Re}\nolimits} \left( {ip_z^{\left( 0 \right)} p_y^{\left( 0 \right)*} } \right)$. The vector ${\bf p}^{(0)}$ is the complex amplitude of the oscillating dipole
\begin{equation}
{\bf{\tilde p}}^{\left( 0 \right)}  = {\mathop{\rm Re}\nolimits} \left[ {{\bf{p}}^{\left( 0 \right)} e^{ - i\omega t} } \right]
\end{equation}
and the tip of this vector moves on an ellipse (polarization ellipse) whose orientation is time independent. Accordingly, the angular vector
\begin{equation}
{\bf{\tilde n}}^{\left( 0 \right)}  = {\bf{\tilde p}}^{\left( 0 \right)}  \times \frac{{d{\bf{\tilde p}}^{\left( 0 \right)} }}{{dt}} = \frac{{i\omega }}{2}{\bf{p}}^{\left( 0 \right)}  \times {\bf{p}}^{\left( 0 \right)*},
\end{equation}
which is orthogonal to the polarization ellipse, is independent on time and the revolution of ${\bf{\tilde p}}^{\left( 0 \right)}$ is counterclockwise with respect to its direction (see panel $\bf a$ of Fig.3). Expanding the angular vector ${\bf{\tilde n}}^{\left( 0 \right)}$ on the cartesian basis we get
\begin{equation}
{\bf{\tilde n}}^{\left( 0 \right)}  = \omega \left[ {{\mathop{\rm Re}\nolimits} \left( { - ip_z^{\left( 0 \right)} p_y^{\left( 0 \right)*} } \right){\bf{\hat e}}_x  + {\mathop{\rm Re}\nolimits} \left( {ip_z^{\left( 0 \right)} p_x^{\left( 0 \right)*} } \right){\bf{\hat e}}_y  + {\mathop{\rm Re}\nolimits} \left( { - ip_x^{\left( 0 \right)*} p_y^{\left( 0 \right)} } \right){\bf{\hat e}}_z } \right]
\end{equation}
which shows that the quantity ${\mathop{\rm Re}\nolimits} \left( {ip_z^{\left( 0 \right)} p_y^{\left( 0 \right)*} } \right)$ is the opposite of its $x$- component (divided by the frequency). As a consequence Eq.(\ref{avJz4}) can be written as
\begin{equation}
\left\langle {J_z } \right\rangle  = 1 - \frac{\gamma }{\omega }\tilde n_x^{\left( 0 \right)}
\end{equation}
which states that the scattered angular momentum exceeds (or is inferior to) that of the impinging plane wave when $0$-th polarization ellipse is negatively (or positively) oriented with respect the $x$-axis. The plain physical meaning of such a statements can be grasped by considering the set of all the $N$ dipoles (as in panel $\bf b$ and $\bf c$ of Fig.3). It is evident that the polarization ellipse of the $q$-th dipole is simply the $0$-th ellipse rotated by an angle $2 \pi q /N$, so that the angular vector ${\bf{\tilde n}}^{\left( q \right)}$ is exactly ${\bf{\tilde n}}^{\left( 0 \right)}$ rotated by the same angle. Therefore the radial component of each vector vector ${\bf{\tilde n}}^{\left( q \right)}$ is equal to $ n_x^{\left( 0 \right)}$ and all the $N$ dipoles rotate on the same direction. Now if $n_x^{\left( 0 \right)} <0$ each dipole at the plane $z=-R$ excites a surface mode on the graphene plane at $z=0$ which preferentially travels in the positive tangential direction (see panel $\bf b$ of Fig.3) \cite{Fortu}; this implies that the overall excited surface mode preferentially rotates counterclockwise and accordingly the scattered angular momentum is positive, in perfect agreement with Eq.(\ref{avJz4}). If $n_x^{\left( 0 \right)} < 0$ the situation is exactly reversed (see panel $\bf c$ of Fig.3) and the scattered angular momentum can also become negative.

\section{Discrete radial emission of plasmons}
The above discussed excitation of modes with definite rotation direction is mainly a consequence of the fact that the dipole polarization ellipse of complex amplitude ${\bf p}^{(0)}$ have a non-vanishing projection onto the tangent $yz$ plane (i.e $p_y^{(0)} \neq 0$ and $p_x^{(0)} \neq 0$). The projection of ${\bf p}^{(0)}$ onto the radial $xz$ plane also plays a role since it is responsible for the excitation of localized surface plasmons radially 
traveling out of the polygon.

In order to discuss this mechanisms we will focus on ${\bf{E}}^{\left( \rm T \right)} \left( {{\bf r}_\bot } \right)$ at resonance, the transmitted field onto the graphene plane $z=0^+$, and we will consider its asymptotical radial behavior for $r_\bot \gg a$. When graphene is resonant, the TE contribution in Eq.(\ref{OAMrep}) can be neglected whereas the TM integral gets its dominant contribution from the wavevectors close to the plasmon wavevector
\begin{equation} \label{kpla}
k_ \bot ^{\left( {\rm pla} \right)}  >  > k_0
\end{equation}
(and, as a consequence we can set $k_{2z}  \cong k_{1z}  \cong ik_ \bot$ in the integrand). Therefore, outside the polygon circle $r_\bot = a$, the condition $k_ \bot r_\bot \gg 1$ is rapidly achieved and Bessel functions there have the asymptotic expansion
\begin{equation} \label{BesselAsym}
J_m \left( {k_ \bot  r_ \bot  } \right) = \sqrt {\frac{2}{{\pi k_ \bot  r_ \bot  }}} \cos \left( {k_ \bot  r_ \bot   - m\frac{\pi }{2} - \frac{\pi }{4}} \right)
\end{equation}
so that, from the second of Eqs.(\ref{UTETM}), we get the asymptotic behavior of the TM modes
\begin{equation} \label{Uasym}
{\bf{U}}_{k_ \bot  ,nN + 1}^{\left( {\rm TM} \right)} \left( {{\bf{r}}_ \bot  } \right) = \frac{1}{{2\pi \sqrt {r_ \bot  } }}\left\{ {e^{ik_ \bot  r_ \bot  } e^{i\left[ {\left( {\varphi  - \frac{\pi }{4}} \right) + nN\left( {\varphi  - \frac{\pi }{2}} \right)} \right]} \left( {{\bf{ e}}_r  + i{\bf{ e}}_z } \right) + e^{ - ik_ \bot  r_ \bot  } e^{i\left[ {\left( {\varphi  + \frac{\pi }{4}} \right) + nN\left( {\varphi  + \frac{\pi }{2}} \right)} \right]} \left( {{\bf{ e}}_r  - i{\bf{ e}}_z } \right)} \right\}.
\end{equation}
where we have introduced the radial cylindrical unit vector ${\bf{e}}_r$ which, together with the azimuthal unit vector ${\bf{e}}_\varphi$ is related to the circular unit vectors by the relations
\begin{eqnarray}
 {\bf{e}}_r  &=& \frac{1}{{\sqrt 2 }}\left( {e^{ - i\varphi } {\bf{e}}_{\rm L}  + e^{i\varphi } {\bf{e}}_{\rm R} } \right), \nonumber \\
 {\bf{e}}_\varphi   &=& \frac{1}{{\sqrt 2 }}\left( { - ie^{ - i\varphi } {\bf{e}}_{\rm L}  + ie^{i\varphi } {\bf{e}}_{\rm R} } \right).
\end{eqnarray}
The TM modes asimptotically split into outgoing ($e^{ik_ \bot  r_ \bot}$) and ingoing ($e^{-ik_ \bot  r_ \bot}$) waves with clockwise $ \left({{\bf{ e}}_r  + i{\bf{ e}}_z } \right)$ and counterclokwise $\left( {{\bf{ e}}_r  - i{\bf{ e}}_z } \right)$ circular polarizations, as seen by the azimuthal unit vector ${\bf{e}}_\varphi$, in the radial plane $rz$. In order to handle the complex amplitudes ${b_{k_ \bot  ,nN + 1}^{\left( {\rm TM} \right)} }$ of these TM modes, note that Eq.(\ref{kpla}) implies that $k_ \bot ^{\left( {\rm pla} \right)} a$ is generally larger than one, so that in the limit $k_ \bot ^{\left( {\rm pla} \right)} a \gg 1$ the Bessel asymptotic behavior of Eq.(\ref{BesselAsym}) with $r_\bot = a$ allows to obtain from the second of Eqs.(\ref{bbb}) the approximate expression
\begin{equation} \label{basym}
b_{k_ \bot  ,nN + 1}^{\left( {\rm TM} \right)}  =  - e^{i\frac{\pi }{4}} \frac{N}{{4\pi \varepsilon _0 \varepsilon _1 }}\frac{{e^{ - k_ \bot  R} k_ \bot  t_{\rm TM} }}{{\sqrt a }}\left[ {\left( {p_z^{\left( 0 \right)}  + ip_x^{\left( 0 \right)} } \right)e^{i\left( {k_ \bot  a - nN\frac{\pi }{2}} \right)}  + i\left( {p_z^{\left( 0 \right)}  - p_x^{\left( 0 \right)} } \right)e^{ - i\left( {k_ \bot  a - nN\frac{\pi }{2}} \right)} } \right].
\end{equation}
Inserting Eqs.(\ref{Uasym}) and (\ref{basym}) into Eq.(\ref{OAMrep}), after neglecting the TE contribution and the incident plane wave contribution, we get
\begin{equation} \label{ETasym}
{\bf{E}}^{\left( \rm T \right)} \left( {{\bf{r}}_ \bot  } \right) = {\bf{E}}_{\rm out}^{\left( \rm T \right)} \left( {{\bf{r}}_ \bot  } \right) + {\bf{E}}_{\rm in}^{\left( \rm T \right)} \left( {{\bf{r}}_ \bot  } \right)
\end{equation}
where the outgoing and ingoing wavefields are
\begin{eqnarray}
{\bf{E}}_{\rm out}^{\left( \rm T \right)} \left( {{\bf{r}}_ \bot  } \right) &=& \left( {{\bf{\hat e}}_r  + i{\bf{\hat e}}_z } \right)\frac{{ - e^{i\varphi } }}{{4\pi \varepsilon _0 \varepsilon _1 \sqrt {r_ \bot  a} }}\left[ {\left( {p_z^{\left( 0 \right)}  + ip_x^{\left( 0 \right)} } \right)I\left( {r_ \bot   + a} \right)\Pi \left( {\varphi  - \pi } \right) + i\left( {p_z^{\left( 0 \right)}  - ip_x^{\left( 0 \right)} } \right)I\left( {r_ \bot   - a} \right)\Pi \left( \varphi  \right)} \right], \nonumber \\
{\bf{E}}_{\rm in}^{\left( \rm T \right)} \left( {{\bf{r}}_ \bot  } \right) &=& \left( {{\bf{\hat e}}_r  - i{\bf{\hat e}}_z } \right)\frac{{e^{i\varphi } }}{{4\pi \varepsilon _0 \varepsilon _1 \sqrt {r_ \bot  a} }}\left[ { - i\left( {p_z^{\left( 0 \right)}  + ip_x^{\left( 0 \right)} } \right)I\left( { - r_ \bot   + a} \right)\Pi \left( \varphi  \right) + \left( {p_z^{\left( 0 \right)}  - ip_x^{\left( 0 \right)} } \right)I\left( { - r_ \bot   - a} \right)\Pi \left( {\varphi  - \pi } \right)} \right], \nonumber \\
\end{eqnarray}
where we have set
\begin{eqnarray}
 I\left( r \right) &=& \int\limits_0^\infty  {dk_ \bot  } k_ \bot  e^{ - k_ \bot  R} t_{\rm TM} e^{ik_ \bot  r}, \nonumber  \\
 \Pi \left( \psi  \right) &=& \frac{N}{{2\pi }}\sum\limits_{n =  - \infty }^{ + \infty } {e^{inN\psi } }.
 \end{eqnarray}
By safely assuming $k_{2z}  \cong k_{1z}  \cong ik_ \bot$ and using the fourth of Eqs.(\ref{rt}), the integral $I\left( r \right)$ can be casted as
\begin{equation}
I\left( r \right) = i\frac{{2\varepsilon _1 k_0 }}{{Z_0 \sigma }}\int\limits_0^\infty  {dk_ \bot  } \frac{{k_ \bot  e^{ - k_ \bot  R} }}{{k_ \bot   - \kappa _ \bot ^{\left( {\rm pla} \right)} }}e^{ik_ \bot  r},
\end{equation}
where the {\it complex} plasmon wave vector is
\begin{equation}
\kappa _ \bot ^{\left( {\rm pla} \right)}  = \frac{{k_0 \left( {\varepsilon _1  + \varepsilon _2 } \right)}}{{\left| {Z_0 \sigma } \right|^2 }}\left[ {{\mathop{\rm Im}\nolimits} \left( {Z_0 \sigma } \right) + i{\mathop{\rm Re}\nolimits} \left( {Z_0 \sigma } \right)} \right].
\end{equation}
\begin{figure*} \label{Fig4}
\center
\includegraphics[width=0.9\textwidth]{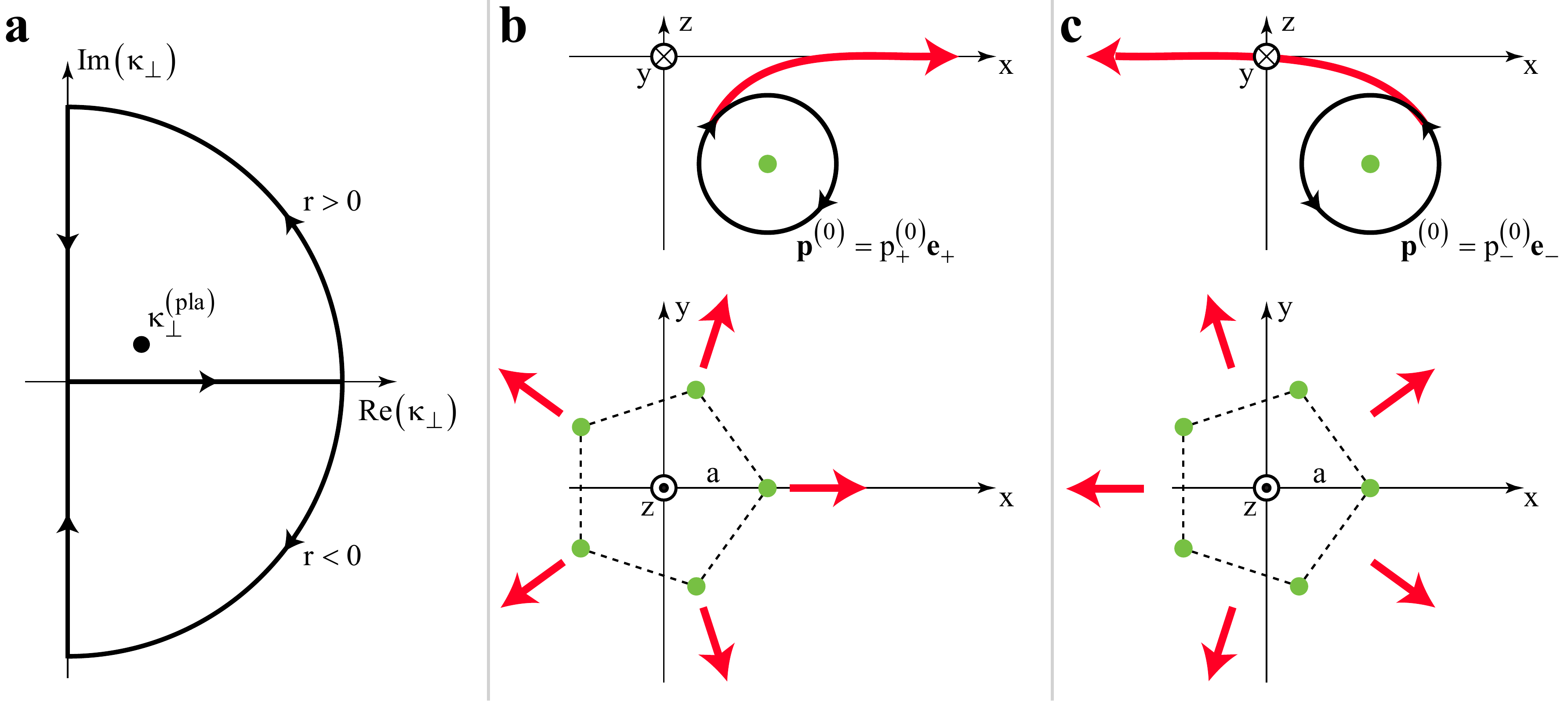}
\caption{(\textbf{a}) Contours in the complex plane $\kappa_\perp$ used to asymptotically evaluate the integral $I (r)$. (\textbf{b}, \textbf{c}) Mechanism of discrete radial emission of plasmons by dipoles positively (\textbf{b}) and negatively (\textbf{c}) circularly polarized in the radial plane $rz$.}
\end{figure*}
Now, resorting to the Jordan lemma, for $r>0$ and $r<0$ the path has to be closed in the upper and lower complex half planes (see Figure 4a), respectively, so that, since ${\mathop{\rm Im}\nolimits} \left[\kappa _ \bot ^{\left( {\rm pla} \right)} \right]  > 0$, we get the contribution of the residue at $\kappa _ \bot ^{\left( {\rm pla} \right)}$ only in the first case. From the residue theorem we get
\begin{equation}
I\left( r \right) = i\frac{{2\varepsilon _1 k_0 }}{{Z_0 \sigma }}\left\{ {\begin{array}{*{20}c} \displaystyle
   { - \int\limits_0^\infty  {d\kappa } \frac{{\kappa e^{ - i\kappa R} }}{{i\kappa  - \kappa _ \bot ^{\left( {\rm pla} \right)} }}e^{ - \kappa r}  + 2\pi i\left[ {\kappa _ \bot ^{\left( {\rm pla} \right)} e^{ - \kappa _ \bot ^{\left( {\rm pla} \right)} R} e^{i\kappa _ \bot ^{\left( {\rm pla} \right)} r} } \right],} & {r > 0,}  \\
  \displaystyle {\int\limits_0^\infty  {d\kappa } \frac{{\kappa e^{i\kappa R} }}{{i\kappa  + \kappa _ \bot ^{\left( {\rm pla} \right)} }}e^{\kappa r} ,} & {r < 0.}  \\
\end{array}} \right.
\end{equation}
Since we are interested in the asymptotic limit $r_\bot \gg a$, it is evident that the two integrals (resulting from the integration along the imaginary axis) appearing in this expression are exponentially small so that, after neglecting them we get
\begin{equation}
I\left( r \right) =  - 2\pi \frac{{2\varepsilon _1 k_0 }}{{Z_0 \sigma }}\kappa _ \bot ^{\left( {\rm pla} \right)} e^{ - \kappa _ \bot ^{\left( {\rm pla} \right)} R} e^{i\kappa _ \bot ^{\left( {\rm pla} \right)} r} \theta \left( r \right)
\end{equation}
Inserting this asymptotic expression into Eq.(\ref{ETasym}), after noting that the theta function kills the contribution of the incoming waves, we get
\begin{eqnarray} \label{ETasymF}
 {\bf{E}}^{\left( \rm T \right)} \left( {{\bf{r}}_ \bot  } \right) = \left( {{\bf{\hat e}}_r  + i{\bf{\hat e}}_z } \right)\frac{{\sqrt 2 k_0 \kappa _ \bot ^{\left( {\rm pla} \right)} e^{ - \kappa _ \bot ^{\left( {\rm pla} \right)} R} }}{{Z_0 \sigma \varepsilon _0 \sqrt {r_ \bot  a} }}e^{i\left( {\kappa _ \bot ^{\left( {\rm pla} \right)} r_ \bot   + \varphi } \right)} \left[ {ip_ + ^{\left( 0 \right)} \sum\limits_{q =  - \infty }^{ + \infty } {\delta \left( {\varphi  - \frac{{2\pi }}{N}q} \right)} e^{ - i\kappa _ \bot ^{\left( {\rm pla} \right)} a} } \right. \nonumber \\
 + \left. {p_ - ^{\left( 0 \right)} \sum\limits_{q =  - \infty }^{ + \infty } {\delta \left( {\varphi  - \frac{{2\pi }}{N}q - \pi } \right)} e^{i\kappa _ \bot ^{\left( {\rm pla} \right)} a} } \right],
\end{eqnarray}
where we have used the relation
\begin{equation}
\Pi \left( \psi  \right) = \frac{N}{{2\pi }}\sum\limits_{n =  - \infty }^{ + \infty } {e^{inN\psi } }  = \sum\limits_{q =  - \infty }^{ + \infty } {\delta \left( {\psi  - \frac{{2\pi }}{N}q} \right)}
\end{equation}
for the Dirac delta comb and we have set
\begin{eqnarray} \label{ppm}
 p_ + ^{\left( 0 \right)}  &=& \frac{1}{{\sqrt 2 }}\left( {p_z^{\left( 0 \right)}  - ip_x^{\left( 0 \right)} } \right), \nonumber \\
 p_ - ^{\left( 0 \right)}  &=& \frac{1}{{\sqrt 2 }}\left( {p_z^{\left( 0 \right)}  + ip_x^{\left( 0 \right)} } \right).
\end{eqnarray}
Equation (\ref{ETasymF}) is the expression of the transmitted field for $r_\bot \gg a$ and it is the main result of this section. This asymptotic field is clockwise circular polarized in the radial $rz$ plane and it is a purely outgoing field whose amplitudes radially fades as $e^{ - {\mathop{\rm Im}\nolimits} \left[ {\kappa _ \bot ^{\left( {\rm pla} \right)} } \right]r_ \bot  } /\sqrt {r_ \bot  }$. As far as its angular dependence, the field is always zero except for the angles
\begin{eqnarray}
 \varphi  &=& \frac{{2\pi }}{N}q , \nonumber \\
 \varphi  &=& \frac{{2\pi }}{N}q + \pi,
\end{eqnarray}
which are respectively excited by the circular $\left( p_x^{\left( 0 \right)}  - ip_z^{\left( 0 \right)} \right)$ and $\left( p_x^{\left( 0 \right)}  + ip_z^{\left( 0 \right)} \right)$ components of the the dipole ${\bf p}^{(0)}$ onto the radial $xz$ plane. We conclude that outside the circle $r_\bot = a$, excitation of radially propagating plasmons occurs and such emission is discrete in the sense that the fields is only radiated along the lines joining the nanospheres and the polygon center. Note that we have obtained the Dirac delta functions in our derivation as a consequence of the condition $k_\perp a \ll 1$ which simplifies the mathematical treatment but it is too much stringent. In actual situations, the radial plasmons do have finite width and amplitude but still they mainly propagate along the radial directions with radial circular polarization as predicted in this section.

The just discussed discrete radial emission of plasmons also admits a simple physical interpretation which is based on the unidirectional excitation a surface mode of Ref.\cite{Fortu}. As a matter of fact, the dipole ${\bf p}^{(0)}$ can be conveniently casted as
\begin{equation}
{\bf{p}}^{\left( 0 \right)}  = p_x^{\left( 0 \right)} {\bf{e}}_x  + p_y^{\left( 0 \right)} {\bf{e}}_y  + p_z^{\left( 0 \right)} {\bf{e}}_z  = p_ + ^{\left( 0 \right)} {\bf{e}}_ +   + p_ - ^{\left( 0 \right)} {\bf{e}}_ -   + p_y^{\left( 0 \right)} {\bf{e}}_y
\end{equation}
where
$p_ + ^{\left( 0 \right)}$ and $p_ - ^{\left( 0 \right)}$ are defined in Eqs.(\ref{ppm}) and
\begin{eqnarray}
 {\bf{e}}_ +   &=& \frac{1}{{\sqrt 2 }}\left( {{\bf{e}}_z  + i{\bf{e}}_x } \right), \nonumber \\
 {\bf{e}}_ -   &=& \frac{1}{{\sqrt 2 }}\left( {{\bf{e}}_z  - i{\bf{e}}_x } \right).
\end{eqnarray}
Now, dipoles purely polarized as ${\bf{e}}_ +$ and as ${\bf{e}}_ -$ circularly spin around the $y-$ axis which sees their rotation in the positive and negative verse, respectively (see Figure 4b and 4c). In the first case we have $p_ + ^{\left( 0 \right)}  \ne 0$ and $p_ - ^{\left( 0 \right)}  = 0$ so that Eq.(\ref{ETasymF}) predicts that plasmons asymptotically travel along the N rays $\varphi = 2 \pi q/N$ which departs from the origin towards the the nano-spheres positions ${\bf{r}}^{\left( q \right)}$ (see Figure 4b). In the second case the situation is reversed, Eq.(\ref{ETasymF}) predicts plasmon emission along the N rays $\varphi = 2 \pi q/N + \pi$ which departs from the origin in the direction opposite to the nano-spheres positions ${\bf{r}}^{\left( q \right)}$ (see Figure 4c). These results are in perfect agreement with the mechanism discussed in Ref.\cite{Fortu} and therefore we conclude that in the region $r_\bot \gg a$ each nanosphere entails the unidirectional excitation of its own surface mode. It is evident that such discrete excitation of radial plasmon is not observed inside the central region $r_\bot \simeq a$ where the effective interference of such modes yields the excitation of modes azimuthally rotating around the $z$ axis, as discussed in the above section.